\documentclass[5p,times]{elsarticle}

\usepackage{graphicx,amsmath,amssymb,booktabs,siunitx,url,doi}
\usepackage{hyperref}
\usepackage{cleveref}
\usepackage{listings}
\usepackage{lineno}
\usepackage{microtype}

\DeclareSIUnit{\cpu}{cpu}

\begin{document}

\begin{frontmatter}

\title{Reclaiming Idle CPU Cycles on Kubernetes: Sparse-Domain Multiplexing\\
for Concurrent MPI-CFD Simulations}

\author[purdue]{Tianfang Xie\corref{cor1}}
\ead{tianfangxie@purdue.edu}
\cortext[cor1]{Corresponding author.}
\affiliation[purdue]{organization={School of Aeronautics and Astronautics, Purdue University},
  city={West Lafayette},
  state={IN},
  postcode={47907},
  country={USA}}

\begin{abstract}
When simulations parallelized with the Message Passing Interface
(MPI) run on shared Kubernetes clusters, conventional CPU scheduling
leaves most provisioned cycles idle at synchronization barriers.
This paper presents a multiplexing framework that reclaims this idle
capacity by co-locating multiple simulations on the same cluster.
Profiling through the MPI profiling interface quantifies per-rank
idle fractions; proportional CPU allocation then lets a second
simulation run concurrently with minimal overhead, yielding
$1.77\times$ throughput.  A sweep to eight concurrent simulations
locates the capacity boundary: throughput rises to $4.09\times$ at
$N=6$, where pod count equals cluster vCPU count, then collapses to
$0.50\times$ at $N=8$ as busy-wait MPI progress threads displace
useful computation.  A single-parameter analytical model fitted on
one mesh reproduces all twelve concurrent configurations, measured
across a $4\times$ range of mesh sizes, within 6.4\%, including the
boundary point within 0.8\%.  Four NAS Parallel Benchmarks kernels
($1.92$--$1.99\times$ at $N=2$), a half-size cluster
($1.68\times$), and per-pod memory accounting confirm generality
across workloads, cluster sizes, and resource dimensions.  A dynamic
controller automates profiling, In-Place Pod Vertical Scaling
(KEP-1287), packing, and fairness monitoring, achieving
$3.25\times$ throughput for four simulations with zero pod restarts,
the first CPU application of in-place resize to running MPI
processes.
\end{abstract}


\end{frontmatter}


\section{Introduction}
\label{sec:introduction}

Cloud computing is increasingly adopted for engineering
computational fluid dynamics
(CFD)~\cite{netto2018hpccloud,beltre2019kubernetes,sochat2025usability}.  Teams deploy
simulations on managed HPC services (e.g., AWS ParallelCluster with
SLURM) or on shared Kubernetes clusters that also host web services,
CI/CD pipelines, and ML training.  For the latter group, maintaining
a separate bare-metal HPC cluster is operationally and financially
impractical~\cite{burns2016borg}.  A typical industrial
Reynolds-averaged Navier--Stokes (RANS) workflow
(automotive aerodynamics, HVAC design, chemical reactor optimization)
uses 4--32 MPI ranks on a mesh of several hundred thousand to a few
million cells; these workloads fit naturally on cloud instances, but
face scheduling constraints unique to container orchestration.

Kubernetes was designed for
stateless microservices, not for tightly-coupled MPI workloads where
a single throttled container stalls the entire communicator.
Within domain-decomposed CFD, near-wall subdomains are
compute-intensive while far-field subdomains require only simple
convection stencils.  Under conventional equal-allocation scheduling,
far-field ranks finish early and idle at MPI barriers.  Our
measurements on a 12-node AWS c5.2xlarge cluster show that sparse
ranks spend 95\% of their time waiting at
\texttt{MPI\_Allreduce} barriers, even when allocated dedicated cores.

On cloud infrastructure, this waste has direct economic consequences:
every millicore reserved but idle represents money spent on
computation that never occurs.  The problem is compounded by the cloud
provider vCPU quotas and Kubernetes per-namespace
\texttt{ResourceQuota} limits that cannot be increased on demand.
Meanwhile, engineering CFD workflows routinely require tens to
hundreds of independent simulations for parameter sweeps,
design-space exploration, and uncertainty quantification.  For such
workloads, \emph{cluster throughput} (simulations completed per hour
per dollar) matters far more than the wall-clock time of any single
case.  The question is not how to make one simulation faster, but
how to complete more simulations within a fixed cloud budget.

Xie~\cite{xie2026rankaware} addressed this imbalance for a single
simulation by mapping subdomain cell counts to per-pod CPU requests,
achieving a 20\% wall-clock reduction under requests-only quality of
service (the Burstable QoS class), which avoids the severe CFS
throttling that hard CPU limits cause at MPI barriers.

The proportional allocation framework frees a substantial fraction of the
cluster's CPU budget: sparse-domain ranks that previously reserved
\SI{1000}{\milli\cpu} each may now request as little as
\SI{67}{\milli\cpu}.  This paper asks the natural follow-up question:
\emph{can the freed capacity be reclaimed by co-scheduling a second
simulation's ranks on the same physical cores, thereby increasing cluster
throughput without additional hardware?}

We address this question through six contributions:
\begin{enumerate}
  \item A profiling methodology based on the MPI standard profiling
        interface (PMPI) that measures per-rank CPU duty cycles,
        quantifying the idle capacity available for multiplexing, and
        a complementary cgroup-accounting method for applications that
        cannot be intercepted at the MPI layer.
  \item A Kubernetes multiplexing framework that co-locates up to
        eight concurrent CFD simulations on shared nodes using
        requests-only scheduling, achieving $4.09\times$ throughput at
        $N=6$ with a Pareto knee at $N=3$ (86\% scheduling
        efficiency).
  \item An analytical throughput model with a single fitted contention
        parameter ($\beta$) and an explicit validity domain.  One
        coefficient fitted on one mesh reproduces twelve concurrent
        configurations across a $4\times$ range of mesh sizes within
        6.4\%; calibration from a single $N=2$ run yields
        conservative capacity estimates, under-predicting throughput
        by at most 11\%.
  \item An empirical characterization of the capacity boundary: when
        the worker pod count exceeds the cluster vCPU count
        ($\rho_N > 1$), busy-wait MPI progress threads displace useful
        computation and throughput collapses below the sequential
        baseline ($0.50\times$ at $N=8$).  The boundary location is
        predicted by the model's own load variable, and the penalty
        for crossing it is quantified.
  \item Generality evidence beyond a single workload and cluster:
        four NAS Parallel Benchmarks kernels co-locate at
        $1.94$--$1.99\times$ ($N=2$, class~D), a six-node subset
        reproduces the dual-simulation gain ($1.68\times$ vs.\
        $1.77\times$ on twelve nodes), and per-pod memory accounting
        shows linear growth of 1.1\,GiB per simulation, far from the
        memory capacity of the cluster.
  \item A dynamic controller that automatically profiles, resizes
        CPU requests (via KEP-1287 In-Place Pod Vertical Scaling), and packs
        additional simulations without manual intervention, achieving
        $3.25\times$ throughput for 4~concurrent simulations with
        64~in-place pod resizes and zero pod restarts, the first
        application of in-place CPU scaling to running MPI workloads.
\end{enumerate}

The remainder of this paper is organized as follows.
Section~\ref{sec:background} provides background on Linux CPU
scheduling and domain-decomposition imbalance.
Section~\ref{sec:related-work} positions this work against
co-scheduling, oversubscription, and cloud-HPC literature.
Section~\ref{sec:idle-characterisation} develops the duty-cycle
metric, the throughput model, and its validity domain.
Section~\ref{sec:framework} describes the design and implementation
of the multiplexing framework and the dynamic controller.
Section~\ref{sec:setup} details the experimental setup and
Section~\ref{sec:results} reports results: duty-cycle
characterization, the factorial comparison, the Pareto sweep to the
capacity boundary, mesh-size sensitivity, benchmark generality,
cluster-size sensitivity, memory footprint, and the dynamic
controller evaluation.
Section~\ref{sec:discussion} discusses practical implications.
Section~\ref{sec:conclusions} concludes.

\section{Background}
\label{sec:background}

\subsection{CFS Weight Scheduling and MPI Barrier Semantics}
\label{sec:cfs-mpi}

The Linux Completely Fair Scheduler (CFS), recently succeeded by
the Earliest Eligible Virtual Deadline First scheduler
(EEVDF)~\cite{stoica1996eevdf} in kernel~6.6+, governs CPU time
distribution among containers on Kubernetes nodes.  Two distinct
mechanisms are relevant~\cite{turner2010cfs,lozi2016linux}.

\paragraph{Proportional sharing (requests-only)}
When a pod specifies CPU \emph{requests} without \emph{limits},
Kubernetes sets the cgroup \texttt{cpu.weight} parameter proportionally.
Under contention, CFS distributes CPU time in proportion to these weights;
when the node is idle, any pod may burst to consume all available cycles.
This Burstable QoS class imposes no hard CPU consumption ceiling.

\paragraph{Hard bandwidth control (with limits)}
When CPU \emph{limits} are set, the CFS bandwidth controller enforces a
hard quota per scheduling period.  Xie~\cite{xie2026rankaware}
showed that this causes a $78\times$ wall-clock inflation for
tightly-coupled MPI workloads, because a single throttled rank stalls
all others at the subsequent barrier.  This incompatibility is the
central motivation for the requests-only allocation model used
throughout this work.

\paragraph{MPI bulk-synchronous semantics}
Iterative CFD solvers follow a bulk-synchronous execution model:
each iteration consists of a compute phase followed by a collective
communication (\texttt{MPI\_Allreduce}, \texttt{MPI\_Barrier}).
All ranks must reach the collective before any can proceed.
Hoefler et~al.~\cite{hoefler2010characterizing} showed that even
microsecond-scale OS noise amplifies logarithmically across the
communicator; CFS throttling produces far more severe amplification.

\subsection{Domain Decomposition Load Imbalance in CFD}
\label{sec:decomp-imbalance}

OpenFOAM~\cite{weller1998openfoam} partitions the computational mesh
into subdomains via \texttt{decomposePar}.  Graph-based
partitioners such as Scotch~\cite{pellegrini1996scotch} and
METIS~\cite{karypis1998metis} minimize inter-processor communication
(edge cut) while producing approximately balanced cell counts across
subdomains.
However, equal cell counts do not imply equal computational cost:
near-wall cells in boundary-layer regions involve turbulence-model
evaluations (e.g., $k$--$\omega$ SST) that are substantially more
expensive than far-field cells, which require only convection and
diffusion stencils.

This inherent imbalance means that far-field ranks finish their compute
phase early and idle at the subsequent \texttt{MPI\_Allreduce} barrier,
waiting for near-wall ranks.  The idle fraction of a rank's iteration
time, which we term the \emph{CPU duty cycle}, represents
computational capacity that is provisioned but unused.

Xie~\cite{xie2026rankaware} exploited this imbalance to reduce
single-simulation wall-clock time by mapping cell counts to
per-pod CPU requests.

The present work takes this observation one step further: if far-field
ranks require only a fraction of a core, the remaining capacity can be
reclaimed by co-scheduling additional simulations' ranks on the same
physical cores.

\section{Related Work}
\label{sec:related-work}

\paragraph{HPC co-scheduling}
Ousterhout~\cite{ousterhout1982scheduling} established that communicating
processes must be scheduled simultaneously to avoid cascading wait times.
Feitelson and Rudolph~\cite{feitelson1992gang} formalized this as gang
scheduling, demonstrating order-of-magnitude speedups for
synchronization-heavy parallel jobs.  Feitelson
et~al.~\cite{feitelson1997theory} later classified jobs as rigid,
moldable, or malleable, observing that gang scheduling wastes resources
when jobs have unequal sizes.  Schwiegelshohn and Yahyapour~\cite{schwiegelshohn2000fairness} further
analyzed fairness guarantees in parallel job scheduling, a property we
evaluate in Section~\ref{sec:concurrent-results}.
Our sparse-domain approach addresses this
at sub-job granularity: by creating intentionally lightweight ranks, we
open scheduling ``holes'' that a second job's ranks can fill.

\paragraph{Oversubscription and contention-aware scheduling}
Running more processes than cores has a long history outside
Kubernetes.  Iancu et~al.~\cite{iancu2010oversubscription} studied
oversubscription of multicore processors directly and found that
moderate oversubscription can improve system throughput when
applications have complementary resource demands, but that
synchronization-heavy codes degrade sharply, which is consistent with
the busy-wait collapse we measure beyond the capacity boundary
(Section~\ref{sec:pareto}).  Utrera
et~al.~\cite{utrera2004malleability} made MPI jobs malleable through
folding, time-sharing several ranks of the same job on one core; our
approach instead time-shares ranks of \emph{different} jobs, chosen so
that their duty cycles are complementary.  Zhuravlev
et~al.~\cite{zhuravlev2010contention} classified schedulers that
mitigate shared-resource contention on multicores, and
Rabenseifner et~al.~\cite{rabenseifner2009hybrid} established hybrid
MPI+OpenMP as the standard way to reduce per-node rank counts on
shared-memory nodes.  These lines of work optimize placement or
programming model on dedicated HPC systems; none addresses the
Kubernetes request/limit abstraction, where the scheduler acts on
declared millicore reservations rather than measured behavior, which
is the layer our framework targets.

\paragraph{Kubernetes for HPC}
Beltre et~al.~\cite{beltre2019kubernetes} provided the foundational
evaluation of MPI on Kubernetes, demonstrating near-bare-metal
performance, but assigning an identical CPU to every rank.
Zhou et~al.~\cite{zhou2021container} proposed a hybrid Kubernetes--TORQUE
architecture for HPC clusters, noting that Kubernetes scheduling is
topology-unaware.  Liu and Guitart~\cite{liu2022scanflow} introduced
two-layer MPI scheduling on Kubernetes but maintained uniform CPU
allocation.  Sochat et~al.~\cite{sochat2025usability} conducted the
largest cloud-HPC benchmark to date (28{,}672 CPUs) without considering
per-rank CPU differentiation or CFD workloads.
Verma et~al.~\cite{verma2015borg} described Google's Borg cluster
manager, and Burns et~al.~\cite{burns2016borg} distilled its lessons
for Kubernetes, noting that safe overcommitment requires understanding
per-application resource sensitivity, a principle that directly
motivates our domain-decomposition-aware CPU allocation.
Rzadca et~al.~\cite{rzadca2020autopilot} demonstrated vertical
autoscaling at Google scale (Autopilot), though their approach targets
long-running services rather than batch MPI workloads.
Pe\~{n}a-Monferrer et~al.~\cite{penamonferrer2021cfd} demonstrated a
hybrid HPC-cloud framework for concurrent CFD analysis and
visualization on Kubernetes, though their work focused on
post-processing pipelines rather than solver-level CPU multiplexing.

\paragraph{Elastic MPI}
Medeiros et~al.~\cite{medeiros2024kub} added horizontal elasticity to
MPI on Kubernetes via checkpoint-restart but explicitly deferred vertical
CPU scaling as future work.  Their follow-on work,
ARC-V~\cite{medeiros2025arcv}, is the only prior application of
In-Place Pod Vertical Scaling to HPC, but it addresses memory
provisioning exclusively; CPU vertical scaling and CFD workloads are
outside its scope.

\paragraph{CFD resource management}
Houzeaux et~al.~\cite{houzeaux2022dynamic} proposed runtime elastic
allocation for the Alya CFD solver on SLURM, adjusting resources by
adding or removing entire MPI ranks.  This requires checkpoint-restart
and operates at whole-core granularity, not fractional CPU.
Priedhorsky and Randles~\cite{priedhorsky2017charliecloud} observed
that HPC container runtimes (Charliecloud, Apptainer/Singularity) deliberately
avoid cgroup-based CPU isolation because it harms MPI performance,
confirming from the opposite direction the CFS--MPI incompatibility
that motivates our requests-only approach.
Younge et~al.~\cite{younge2017containers} and
Rudyy et~al.~\cite{rudyy2019containers} independently confirmed that
container overhead is negligible for well-configured HPC workloads,
supporting our choice of containerized deployment.

\paragraph{Memory-bandwidth contention}
Mars et~al.~\cite{mars2011bubbleup} introduced the Bubble-Up methodology
for predicting co-location interference, finding that memory bandwidth,
not CPU, is the dominant source of degradation for memory-intensive
workloads.  Zhang et~al.~\cite{zhang2013cpi2} deployed CPI-based
contention detection at Google scale, confirming that memory-bandwidth
interference is a production concern.
Hager et~al.~\cite{hager2016exploring} showed via the ECM
(Execution-Cache-Memory) model that stencil-based PDE solvers,
including finite-volume CFD, are fundamentally memory-bandwidth-bound.
Lo et~al.~\cite{lo2015heracles} demonstrated that hardware-level
isolation via Intel Cache Allocation Technology (CAT) enables safe
co-location at 90\% utilization.
Delimitrou and Kozyrakis~\cite{delimitrou2014quasar} showed that
collaborative filtering of workload interference signatures can
automate co-location decisions at cluster scale.
These results suggest that memory-bandwidth saturation may become the
practical ceiling for multiplexing as co-location density increases,
although this depends on the per-rank duty cycle of the workload.

\paragraph{Gap addressed by this work}
Table~\ref{tab:rw2} summarizes the positioning.
No prior work has empirically characterized what happens when two
tightly-coupled MPI-CFD simulations share Kubernetes nodes with CFS
weight-based CPU multiplexing on sparse-domain ranks.
No prior work combines all five elements: Kubernetes scheduling,
per-rank fractional CPU allocation, domain-decomposition awareness,
CFD workloads, and multi-simulation co-location.  Queue-level
schedulers such as Volcano and Kueue gang-schedule whole jobs
sequentially when resources are scarce; the sequential execution they
produce is exactly the baseline our concurrent configurations are
measured against, and our rank-level mechanism composes with them
rather than replacing them.

\begin{table}[t]
  \centering
  \caption{Positioning relative to prior work.
    \checkmark\ = yes.}
  \label{tab:rw2}
  \small
  \begin{tabular}{@{}lcccc@{}}
    \toprule
    \textbf{Work} & \textbf{K8s} & \textbf{CFD} &
    \textbf{Prop.\ CPU} & \textbf{Multi-job} \\
    \midrule
    Beltre et al.~\cite{beltre2019kubernetes}    & \checkmark &            &            &            \\
    Houzeaux et al.~\cite{houzeaux2022dynamic}   &            & \checkmark &            &            \\
    Medeiros/ARC-V~\cite{medeiros2025arcv}       & \checkmark &            &            &            \\
    Mars et al.~\cite{mars2011bubbleup}          &            &            &            & \checkmark \\
    Lo et al.~\cite{lo2015heracles}              &            &            &            & \checkmark \\
    Pe\~{n}a-Monferrer~\cite{penamonferrer2021cfd}       & \checkmark & \checkmark &            &            \\
    Iancu et al.~\cite{iancu2010oversubscription} &            &            &            & \checkmark \\
    Zhuravlev et al.~\cite{zhuravlev2010contention} &          &            &            & \checkmark \\
    Xie~\cite{xie2026rankaware}                   & \checkmark & \checkmark & \checkmark &            \\
    \textbf{This work}                           & \checkmark & \checkmark & \checkmark & \checkmark \\
    \bottomrule
  \end{tabular}
\end{table}

\section{CPU Idle Characterization and Throughput Model}
\label{sec:idle-characterisation}

To quantify the CPU idle time available for multiplexing, we instrument
each MPI rank with a lightweight profiling wrapper and define a formal
duty-cycle metric.

\subsection{PMPI Profiling Methodology}
\label{sec:pmpi}

The MPI standard~\cite{mpiforum2021} defines the PMPI (Profiling MPI)
interface, which allows user-supplied wrapper functions to intercept
MPI calls without modifying the application source code.  We implement a shared library,
\texttt{libpmpi\_trace.so}, that interposes on six collective
and synchronization operations:
\texttt{MPI\_Barrier}, \texttt{MPI\_Allreduce},
\texttt{MPI\_Alltoall}, \texttt{MPI\_Sendrecv},
\texttt{MPI\_Wait}, and \texttt{MPI\_Waitall}.
For each intercepted call, the wrapper records the rank index, function
name, and entry/exit timestamps via
\texttt{clock\_gettime(CLOCK\_MONOTONIC)} at microsecond resolution.
Entries are buffered in memory and flushed to per-rank CSV files at
\texttt{MPI\_Finalize}.

The library is loaded at runtime via \texttt{LD\_PRELOAD}, requiring
no recompilation of OpenFOAM or the MPI runtime.

\subsection{Duty-Cycle Definition}
\label{sec:duty-cycle}

We define the \emph{CPU duty cycle} of rank~$i$ at iteration~$k$ as the
fraction of the iteration wall-clock time spent in computation rather
than MPI communication or synchronization:

\begin{equation}
  d_i^{(k)} = \frac{t_{\text{compute},i}^{(k)}}
               {t_{\text{compute},i}^{(k)} + t_{\text{MPI},i}^{(k)}},
  \label{eq:duty-cycle}
\end{equation}

\noindent where $t_{\text{compute}}$ is the time between the exit of one
MPI call and the entry of the next (i.e., the solver sweep), and
$t_{\text{MPI}}$ is the cumulative time spent inside all intercepted MPI
calls during iteration~$k$.  We detect iteration boundaries by counting
\texttt{MPI\_Barrier} calls, which OpenFOAM invokes once at the end of
each solver iteration.

The steady-state duty cycle of rank~$i$ is then the time-weighted average
over all iterations after the initial transient:
\begin{equation}
  \bar{d}_i = \frac{\sum_{k=k_0}^{K} t_{\text{compute},i}^{(k)}}
              {\sum_{k=k_0}^{K}
                \bigl(t_{\text{compute},i}^{(k)} + t_{\text{MPI},i}^{(k)}\bigr)},
  \label{eq:duty-cycle-avg}
\end{equation}
where $k_0$ skips the first 10\% of iterations to exclude startup
effects.

\subsection{Theoretical Multiplexing Capacity}
\label{sec:theory}

If rank~$i$ has a steady-state duty cycle $\bar{d}_i$ and is allocated
$r_i$~millicores of CPU, then the \emph{reclaimable capacity} of that
rank is:
\begin{equation}
  c_i = r_i \cdot (1 - \bar{d}_i).
  \label{eq:reclaimable}
\end{equation}

The aggregate reclaimable capacity across all ranks of a simulation
is $C = \sum_{i=0}^{N-1} c_i$.  Under proportional allocation with
weights $(w_0, \ldots, w_{N-1})$ and a total CPU budget~$B$,
$r_i = B \cdot w_i / \sum_j w_j$.  Sparse-domain ranks (low~$w_i$) with
low duty cycles (low~$\bar{d}_i$) contribute disproportionately to~$C$
because their allocated CPU is mostly idle.

This reclaimable capacity represents the theoretical upper bound on the
CPU budget available for a co-located second simulation.  In practice,
CFS context-switch overhead and MPI collective contention reduce the
effective gain.

\paragraph{Capacity boundary under busy-wait MPI}
The duty cycle measures \emph{useful} computation, but it does not
equal CPU occupancy.  Open~MPI's progress engine busy-polls during
collective waits by default (\texttt{mpi\_yield\_when\_idle=0}), so a
rank idling at a barrier still occupies a full core with spin cycles.
While each rank has a core to itself, this spin is harmless: CFS
preempts spinning threads at negligible cost when a co-located rank
becomes runnable, which is precisely why weight-based multiplexing
works.  The regime changes when the number of runnable threads exceeds
the number of cores.  Spin cycles then compete with useful computation
on every core, and each additional simulation slows all others
superlinearly.  The framework therefore has a hard capacity boundary
at
\begin{equation}
  N_{\max} = \left\lfloor \frac{MC}{R} \right\rfloor,
  \label{eq:capacity-boundary}
\end{equation}
where $M$ is the number of worker nodes, $C$ the vCPUs per node,
and $R$ the ranks per simulation: the largest $N$ for which every
worker pod can occupy its own vCPU ($\rho_{N} \le 1$ in the
notation of Section~\ref{sec:model}).
Section~\ref{sec:pareto} confirms this
boundary empirically: on our cluster ($M{=}12$, $C{=}8$, $R{=}16$,
$N_{\max}{=}6$), throughput peaks at $N=6$ and collapses beyond it.

\subsection{Throughput Prediction Model}
\label{sec:model}

We model the makespan of $N$ concurrent simulations sharing a cluster
of $M$ nodes with $C$~vCPUs each.  Define the \emph{normalized cluster
load} $\rho_N = NR/(MC)$, where $R$ is the number of ranks per
simulation.  When $\rho_N \ll 1$, pods rarely compete for CPU; as
$\rho_N \to 1$, every vCPU hosts a pod, and context-switch overhead
grows.  We propose a linear contention model:
\begin{equation}
  T_N = T_1 \bigl(1 + \beta\,(\rho_N - \rho_1)\bigr),
  \label{eq:makespan-model}
\end{equation}
where $T_1$ is the single-simulation makespan, $\rho_1 = R/(MC)$ is
the baseline load, and $\beta$ is a \emph{contention coefficient} that
captures the per-unit-load overhead from CFS scheduling and TCP
contention for MPI collectives.

The throughput gain follows directly:
\begin{equation}
  \Theta(N) = \frac{N\,T_1}{T_N}
            = \frac{N}{1 + \beta\,(\rho_N - \rho_1)}.
  \label{eq:throughput-model}
\end{equation}

This single-parameter model has a practical advantage: $\beta$ can be
fitted from a single dual-simulation experiment ($N=2$), then used to
predict throughput for any~$N$ within the validity domain
$\rho_N \le 1$ established by
Equation~\ref{eq:capacity-boundary}.  Beyond that boundary the
linear-contention assumption fails by construction, because spin
cycles displace computation rather than merely adding scheduling
overhead.  The model makes no claim outside its domain; instead, the
domain edge itself is a prediction, which
Section~\ref{sec:pareto} tests directly.
Because $\beta$ absorbs workload-dependent contention (cache pressure,
collective sizes), it is a per-workload-class constant rather than a
universal one.  Section~\ref{sec:beta-sensitivity} quantifies how far
a single fitted $\beta$ transfers across mesh resolutions.

\section{Design and Implementation}
\label{sec:framework}

\begin{figure}[t]
  \centering
  \includegraphics[width=\linewidth]{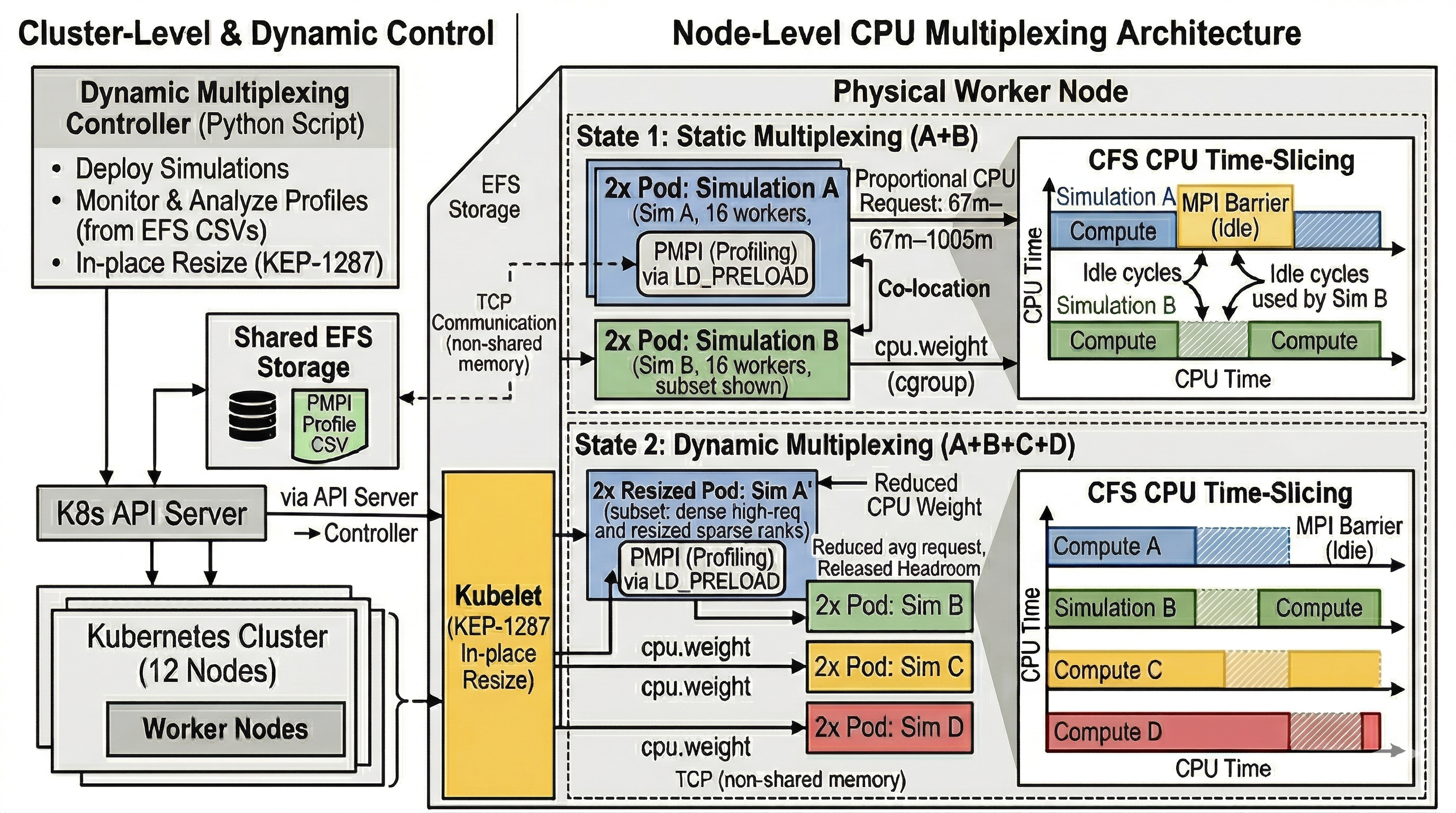}
  \caption{System architecture.  Left: cluster-level control flow
    showing the dynamic controller, K8s API server, and shared EFS
    storage.  Right: node-level CPU multiplexing in two states.
    State~1 (static): two co-located simulations share CPU via CFS
    weight-based time-slicing; idle MPI barrier cycles from one
    simulation are reclaimed by the other.
    State~2 (dynamic): the controller resizes Sim~A's CPU requests
    via KEP-1287 (kubelet in-place resize), releasing headroom for
    Sims~B--D.}
  \label{fig:architecture}
\end{figure}

This section describes the design of the multiplexing framework and
the Kubernetes-native mechanisms it is built on.
Figure~\ref{fig:architecture} provides a high-level overview:
multiple simulations share worker nodes via CFS weight-based
scheduling, a PMPI profiler measures per-rank duty cycles, and a
dynamic controller automates resizing and packing.

\paragraph{Design goals}
The framework is built around three explicit goals.
First, \emph{no application changes}: profiling attaches through
\texttt{LD\_PRELOAD} and scheduling acts only through standard pod
resource fields, so any MPI application runs unmodified.
Second, \emph{no scheduler changes}: co-location emerges from
declarative CPU requests interpreted by the stock Kubernetes
scheduler and the Linux CFS, not from a custom scheduler plugin,
which keeps the mechanism deployable on managed clusters where
administrators cannot replace system components.
Third, \emph{graceful composition}: the mechanism operates at the
rank level within one namespace and composes with queue-level gang
schedulers (Volcano, Kueue) that decide which jobs run at all.
The novelty is not any individual mechanism, all of which are stock
Kubernetes features, but the demonstration that duty-cycle-derived
fractional requests turn the CFS weight system into an effective
multiplexer for tightly-coupled MPI workloads, plus the
quantification of exactly how far this can be pushed
(Section~\ref{sec:pareto}).

\subsection{Pod Deployment and CPU Allocation}
\label{sec:k8s-implementation}

Each simulation is deployed as a set of 16 worker pods (one per MPI
rank) plus a dedicated launcher pod that executes \texttt{mpirun}.
For two concurrent simulations~A and~B, the cluster hosts 34~pods.  Simulation pods are distinguished by a \texttt{sim} label
(\texttt{sim=A} or \texttt{sim=B}) and use separate case directories
on the shared Amazon Elastic File System (EFS) volume to prevent
solver output conflicts.
Each launcher pod receives its own MPI hostfile (deployed as a
Kubernetes ConfigMap) containing only the pod IPs of its simulation.

All worker pods specify CPU \emph{requests} only, with no
\emph{limits} (Burstable QoS), for the reasons established in
Section~\ref{sec:cfs-mpi}: hard limits cause CFS bandwidth
throttling at MPI barriers.

When two simulations share the cluster, the Kubernetes scheduler
interleaves their pods across the same nodes.  We do not use pod
affinity/anti-affinity rules, as the default bin-packing behavior
distributes pods across nodes.  Section~\ref{sec:setup} details
the cluster hardware.

\subsection{MPI Transport Isolation}
\label{sec:mpi-isolation}

When two independent MPI jobs share a physical node, their processes
may inadvertently discover each other through the shared-memory byte
transfer layer (Open~MPI's \texttt{vader} BTL).  Although separate
\texttt{MPI\_COMM\_WORLD} communicators prevent logical interference,
the shared-memory transport path can introduce memory-bus contention
that is difficult to attribute.  To ensure clean isolation between
simulations, we disable the shared-memory BTL and restrict each job
to TCP transport (\texttt{-{}-mca btl tcp,self}).  This eliminates
cross-job shared-memory interference at the cost of slightly higher
intra-node latency, which is acceptable for the coarse-grained
collective operations in OpenFOAM's iterative solver.

\subsection{Pod Specification}
\label{sec:pod-spec}

Listing~\ref{lst:pod-manifest} shows a representative pod
specification for a sparse-domain rank of simulation~A.
\begin{lstlisting}[
  caption={Pod specification for a sparse rank (67m CPU request).},
  label={lst:pod-manifest}]
apiVersion: v1
kind: Pod
metadata:
  name: of-worker-a-0
  labels:
    app: openfoam
    role: worker
    sim: A
spec:
  containers:
  - name: openfoam
    image: openfoam-k8s:v10
    resources:
      requests:
        cpu: "67m"
      # No limits: Burstable QoS
    resizePolicy:
    - resourceName: cpu
      restartPolicy: NotRequired
\end{lstlisting}

Walking through Listing~\ref{lst:pod-manifest} line by line: the
\texttt{sim: A} label (line~8) partitions pods into per-simulation
groups that the launcher's hostfile and the metrics pipeline select
on; the \texttt{requests.cpu: "67m"} field (line~15) is the entire
scheduling interface of the framework, as this single number becomes
the pod's CFS \texttt{cpu.weight} and thereby its share of a
contended core; the deliberate absence of a \texttt{limits} block
(line~16) keeps the pod in the Burstable QoS class so that idle
cycles from other pods remain harvestable; and the
\texttt{resizePolicy} block (lines~17--19) declares that CPU changes
apply in place, without container restart, which is the hook the
dynamic controller uses.  The value \texttt{67m} itself is not
hand-chosen: it is the duty-cycle-proportional share of the sparse
ranks under the 1:5:15 weight vector described in
Section~\ref{sec:setup}.

The \texttt{resizePolicy} field enables In-Place Pod Vertical Scaling
(KEP-1287~\cite{kep1287}), permitting runtime CPU adjustments without
pod restart.

\subsection{Duty-Cycle Measurement Without PMPI}
\label{sec:cgroup-method}

The PMPI wrapper intercepts the C bindings of MPI.  Applications that
call MPI through Fortran interfaces, or binaries that cannot be
preloaded, require an alternative signal.  For these cases the
framework snapshots the cgroup~v2 CPU accounting file
(\texttt{cpu.stat}, field \texttt{usage\_usec}) of every worker pod
immediately before and after a run; the difference divided by the
elapsed wall-clock time yields the per-pod CPU occupancy.  Occupancy
is an upper bound on the duty cycle because busy-wait spin counts as
CPU usage (Section~\ref{sec:theory}); Section~\ref{sec:duty-results}
quantifies the gap on instrumented runs where both signals are
available.  The occupancy signal is the right input for capacity
planning, since spin occupies cores regardless of its usefulness,
and it is the signal we use for the NAS Parallel Benchmarks in
Section~\ref{sec:npb-results}.

\subsection{Dynamic Controller Design}
\label{sec:controller-design}

The dynamic controller automates the full multiplexing pipeline as a
four-phase loop implemented in roughly 500~lines of Python against
the Kubernetes API:
\begin{description}
  \item[Phase A (Profile).] Attach the PMPI wrapper to a running
    simulation for a 50-iteration window and compute per-rank duty
    cycles $\bar{d}_i$.
  \item[Phase B (Resize).] Set each rank's CPU request to
    $r_i = r_{\text{base}} \cdot \bar{d}_i / \max_j \bar{d}_j$ and
    apply all 16 changes through KEP-1287 in-place resize; the MPI
    processes keep running throughout.
  \item[Phase C (Pack).] Compare aggregate requests against
    allocatable cluster CPU; while headroom exceeds one simulation's
    footprint, deploy the next simulation, then profile and resize it
    in turn.
  \item[Phase D (Monitor).] Track per-simulation progress rates from
    solver logs; if the fastest-to-slowest ratio exceeds a fairness
    threshold, apply one bounded weight adjustment (at most two per
    simulation, preventing oscillation).
\end{description}
Phases A and B embody the measurement-driven principle of the
framework: requests follow measured duty cycles rather than static
guesses.  Phase C enforces the capacity boundary of
Equation~\ref{eq:capacity-boundary} by construction, since it stops
packing when requests approach allocatable capacity.
Section~\ref{sec:dynamic-eval} evaluates the controller end to end.


\section{Experimental Setup}
\label{sec:setup}

All experiments run on Amazon EC2 c5.2xlarge instances, each equipped
with an Intel Xeon Platinum 8124M processor at \SI{3.0}{\giga\hertz}
(Skylake/Cascade Lake microarchitecture), 4~physical cores with
hyper-threading (8~vCPUs), \SI{16}{\gibi\byte} RAM, and a single
non-uniform memory access (NUMA) domain.  The c5 family provides non-burstable, compute-optimized
instances with dedicated CPU resources.  The use of non-burstable instances is critical:
Xie~\cite{xie2026rankaware} observed that burstable t3 instances
produce irreproducible benchmark results due to CPU credit
variability (identical configurations yielded 563\,s vs.\ 236\,s
across sessions).  Table~\ref{tab:hardware} summarizes the cluster
composition.

\begin{table}[t]
  \centering
  \caption{Cluster composition (AWS EC2, 98~vCPU total).}
  \label{tab:hardware}
  \small
  \begin{tabular}{@{}lccccc@{}}
    \toprule
    \textbf{Instance} & \textbf{Count} & \textbf{vCPU} & \textbf{RAM} &
    \textbf{NUMA} & \textbf{Role} \\
    \midrule
    t3.medium  & 1  & 2 & \SI{4}{\gibi\byte}  & 1 & Control plane \\
    c5.2xlarge & 12 & 8 & \SI{16}{\gibi\byte} & 1 & Worker \\
    \midrule
    \multicolumn{2}{@{}l}{\textbf{Total workers}} & \multicolumn{1}{c}{\textbf{96}} & & & 12 nodes \\
    \bottomrule
  \end{tabular}
\end{table}

A 16-rank simulation places 1--2 ranks per node
(16~ranks across 12~worker nodes); a dual-simulation experiment
places 2--3 ranks per node.
The cluster is managed by k3s~v1.35.0+k3s1, a lightweight Kubernetes distribution
that supports the In-Place Pod Vertical Scaling feature (generally
available since Kubernetes~v1.33).  The control-plane node is tainted
\texttt{NoSchedule} to prevent workload pods from being placed on it.
All worker nodes mount a shared Amazon EFS volume for case files
and solver output.

On the software side, each worker pod runs a custom Docker image
containing OpenFOAM~10 (Foundation release), Open~MPI~4.1, and an
SSH daemon for MPI inter-pod communication.  The image is publicly
available (see Data Availability).  The PMPI profiling wrapper
(\texttt{libpmpi\_trace.so}) is compiled on the cluster and volume-mounted
into each pod; no modification to the OpenFOAM binary is required.

The benchmark case is a NACA~0012 aerofoil at Mach 0.72, solved with
\texttt{rhoSimpleFoam} (compressible steady-state) with
the $k$--$\omega$ SST turbulence model.  The mesh contains
498\,834 hexahedral cells, generated by scaling the OpenFOAM tutorial
\texttt{blockMeshDict} parameters.  Boundary conditions: freestream
velocity $U = \SI{250}{\metre\per\second}$, temperature
$T = \SI{298}{\kelvin}$, pressure $p = \SI{1e5}{\pascal}$, angle of
attack $\alpha = 0^{\circ}$.

Domain decomposition uses a manual distance-based partitioner that
assigns cells to 16~processors in three concentric zones based on
each cell's distance from the aerofoil center ($x/c = 0.5$).
The weight vector ${\small(1, 1, 1, 1, 1, 1, 1, 1},$ ${\small 5, 5, 5, 5, 15, 15, 15, 15)}$
determines the cell fraction per zone:
the four dense ranks (weight~15, ranks~12--15) receive the innermost
68\% of cells (boundary layer and near-wall region),
the four medium ranks (weight~5, ranks~8--11) receive the next 23\%,
and the eight sparse ranks (weight~1, ranks~0--7) receive the
outermost 9\% (far-field).
Within each zone, cells are subdivided into equal angular sectors.
This concentric layout ensures that sparse ranks contain exclusively
far-field cells with low computational cost per cell, maximizing the
idle fraction available for multiplexing.
Xie~\cite{xie2026rankaware} demonstrated the effectiveness of
concentric decomposition at 4~ranks, where it reduced wall-clock
time by 19\% compared to equal Scotch partitioning.
The present work extends this approach to 16~ranks via an automated distance-sorting script (see Data Availability).

\begin{figure}[t]
  \centering
  \includegraphics[width=\linewidth]{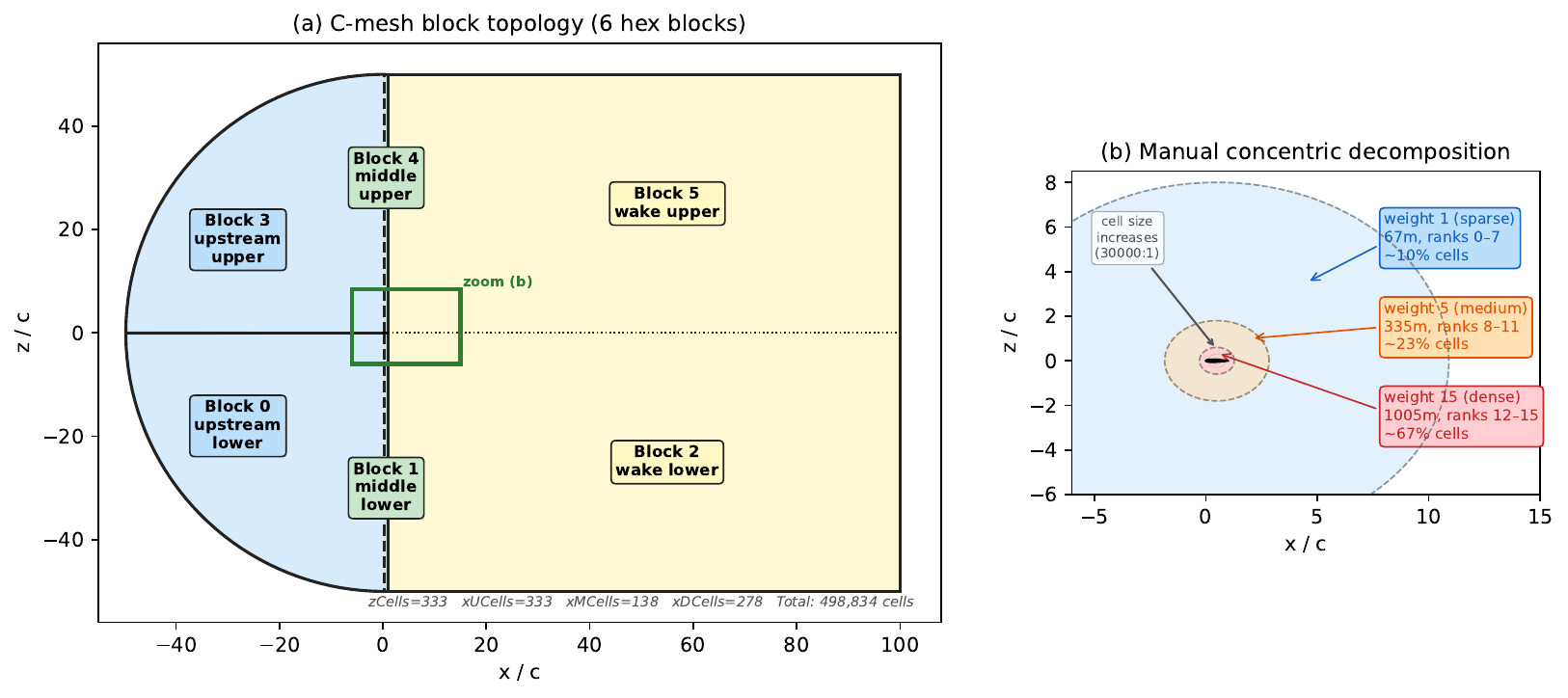}
  \caption{(a)~C-mesh block topology (6~hex blocks, 498\,834~cells).
    (b)~Concentric weight zones for 16-rank decomposition: dense
    ranks (weight~15) receive near-wall cells, sparse ranks (weight~1)
    receive far-field cells.}
  \label{fig:mesh-blocks}
\end{figure}

\begin{figure}[t]
  \centering
  \includegraphics[width=\linewidth]{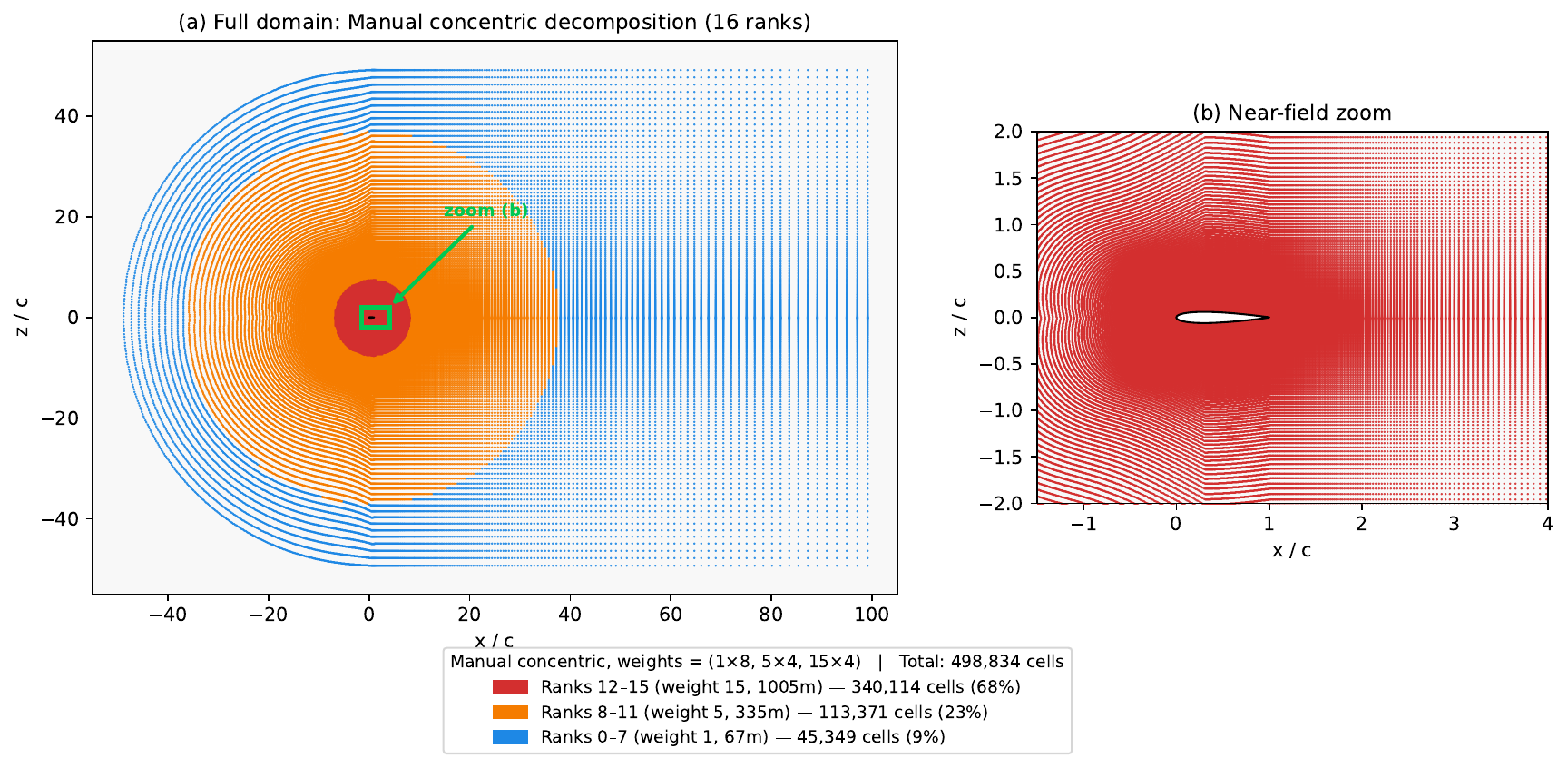}
  \caption{Manual concentric decomposition of the NACA~0012 mesh into
    16~ranks.  Cell centers are colored by processor assignment.
    Dense ranks (red, 68\% of cells) occupy the near-wall region,
    medium ranks (orange, 23\%) the intermediate zone, and sparse ranks
    (blue, 9\%) the far-field.}
  \label{fig:decomposition}
\end{figure}

Figure~\ref{fig:mesh-blocks} shows the C-mesh block topology and the
concentric weight zones.  Figure~\ref{fig:decomposition} visualizes the
resulting cell-to-processor assignment: the innermost ring (red) contains
the dense near-wall cells assigned to weight-15 ranks, while the outermost
ring (blue) contains the sparse far-field cells assigned to weight-1 ranks.

Table~\ref{tab:configs} lists the experimental configurations, which
follow a $2 \times 2$ factorial design with two factors: CPU allocation
strategy (equal vs.\ proportional) and co-location density (single
simulation vs.\ dual simulation).

\begin{table}[t]
  \centering
  \caption{Experimental configurations ($2 \times 2$ factorial).
    CPU values are per-rank requests (millicores); no limits are set.}
  \label{tab:configs}
  \small
  \begin{tabular}{@{}llccl@{}}
    \toprule
    \textbf{Config} & \textbf{Sims} & \textbf{Ranks} &
    \textbf{CPU/rank} & \textbf{QoS} \\
    \midrule
    C-1E & 1 & 16 & 1000m (equal)    & Burstable \\
    C-1P & 1 & 16 & 67--1005m (prop.)   & Burstable \\
    C-2E & 2 & $2 \times 16$ & 1000m (equal)  & Burstable \\
    C-2P & 2 & $2 \times 16$ & 67--1005m (prop.) & Burstable \\
    C-$N$P & 3--8 & $N \times 16$ & 67--1005m (prop.) & Burstable \\
    \bottomrule
  \end{tabular}
\end{table}

The Pareto sweep of Section~\ref{sec:pareto} extends the
proportional configuration to $N = 3\ldots8$ (rows C-$N$P); each
added simulation contributes 5.9\,vCPU of requests, so aggregate
requests grow from 17.7\,vCPU at $N=3$ to 47.2\,vCPU at $N=8$,
always within the 96-vCPU allocatable capacity.

Configurations C-1E and C-1P serve as isolated baselines on the
c5.2xlarge cluster (498\,834-cell mesh), analogous to the 4-rank
experiments in Xie~\cite{xie2026rankaware} but at 16~ranks and
$5\times$ larger mesh.

All configurations run on the same 98-vCPU cluster
(\SI{96}{vCPU} of worker capacity across 12~homogeneous nodes),
ensuring a fair comparison where the only variable is the CPU
allocation strategy.

Configuration~C-2E requires $2 \times 16 \times
\SI{1000}{\milli\cpu} = \SI{32}{vCPU}$ of aggregate requests,
consuming one-third of the cluster's \SI{96}{vCPU} capacity.
Configuration~C-2P requires only
$2 \times \SI{5900}{\milli\cpu} \approx \SI{12}{vCPU}$
of aggregate requests, leaving \SI{84}{vCPU} of headroom.
The gap between C-2E's \SI{32}{vCPU} and C-2P's \SI{12}{vCPU}
represents the CPU budget freed by proportional allocation.
Both configurations have access to the same physical hardware;
the difference lies solely in how much CPU each rank \emph{reserves}
via its Kubernetes request, which determines the CFS scheduling
weight when cores are contended.

\paragraph{Mesh-size variants}
To test whether the throughput model transfers across workload
intensity, two additional meshes are generated by rescaling the same
\texttt{blockMeshDict}: a 249\,100-cell variant (250K) and a
997\,578-cell variant (1M), spanning a $4\times$ range around the
498\,834-cell baseline (500K).  Each variant is decomposed with the
same concentric 1:5:15 weight vector and run under configurations
C-1P, C-2P, and C-3P.

\paragraph{Second benchmark family}
To test generality beyond CFD, we run four kernels of the NAS
Parallel Benchmarks (NPB~3.4.3, MPI flavor, class~D, 16~ranks):
EP (embarrassingly parallel), CG (conjugate gradient,
communication-intensive), LU (lower-upper Gauss-Seidel), and BT
(block tridiagonal).  NPB kernels are load-balanced by design, so
proportional sparse-domain allocation does not apply; instead, all
pods receive equal 1000m requests, and the experiment probes
co-location overhead at $N{=}1$ versus $N{=}2$ across a wide range of
communication intensities.  Because the kernels call MPI through
Fortran bindings, duty information comes from the cgroup occupancy
method of Section~\ref{sec:cgroup-method}.

\paragraph{Cluster-size subset}
To test sensitivity to cluster size, configurations C-1P and C-2P are
repeated with six of the twelve worker nodes cordoned, halving
capacity to 48~vCPUs and doubling per-node pod density.

\paragraph{Repetitions and statistics}
Every configuration is repeated three times, with one exception:
C-8P, whose runs each take 5.5~hours in the over-capacity regime, is
repeated twice; the three C-7P repetitions in the same regime vary by
under 2\%, supporting the reduced count.  We report mean $\pm$
standard deviation.  Run-to-run coefficients of variation are below
2.4\% for every configuration (below 1.1\% for all $N \ge 2$), and
per-simulation fairness (slowest-to-fastest wall-clock ratio within a
run) stays below 1.08 for every concurrent configuration up to
$N=8$.  No run failed or was discarded; solver logs confirm all runs
completed the full 200 iterations (100 for the 1M mesh) with the
standard convergence behavior.

The primary metric is \emph{wall-clock time} (seconds), extracted
from the OpenFOAM solver log.  From this we derive \emph{throughput}
(completed simulations divided by the elapsed wall-clock duration)
and \emph{per-case degradation}
$(T_{\text{concurrent}} - T_{\text{isolated}}) /
T_{\text{isolated}} \times 100$\%.
Node CPU utilization is sampled every \SI{5}{\second} to monitor
resource contention.  Per-rank duty cycles are obtained from the
PMPI profiling wrapper described in Section~\ref{sec:pmpi};
per-pod memory is sampled every \SI{5}{\second} from the Kubernetes
metrics API.

\section{Results}
\label{sec:results}

\subsection{Duty-Cycle Characterization}
\label{sec:duty-results}

\begin{figure}[t]
  \centering
  \includegraphics[width=\linewidth]{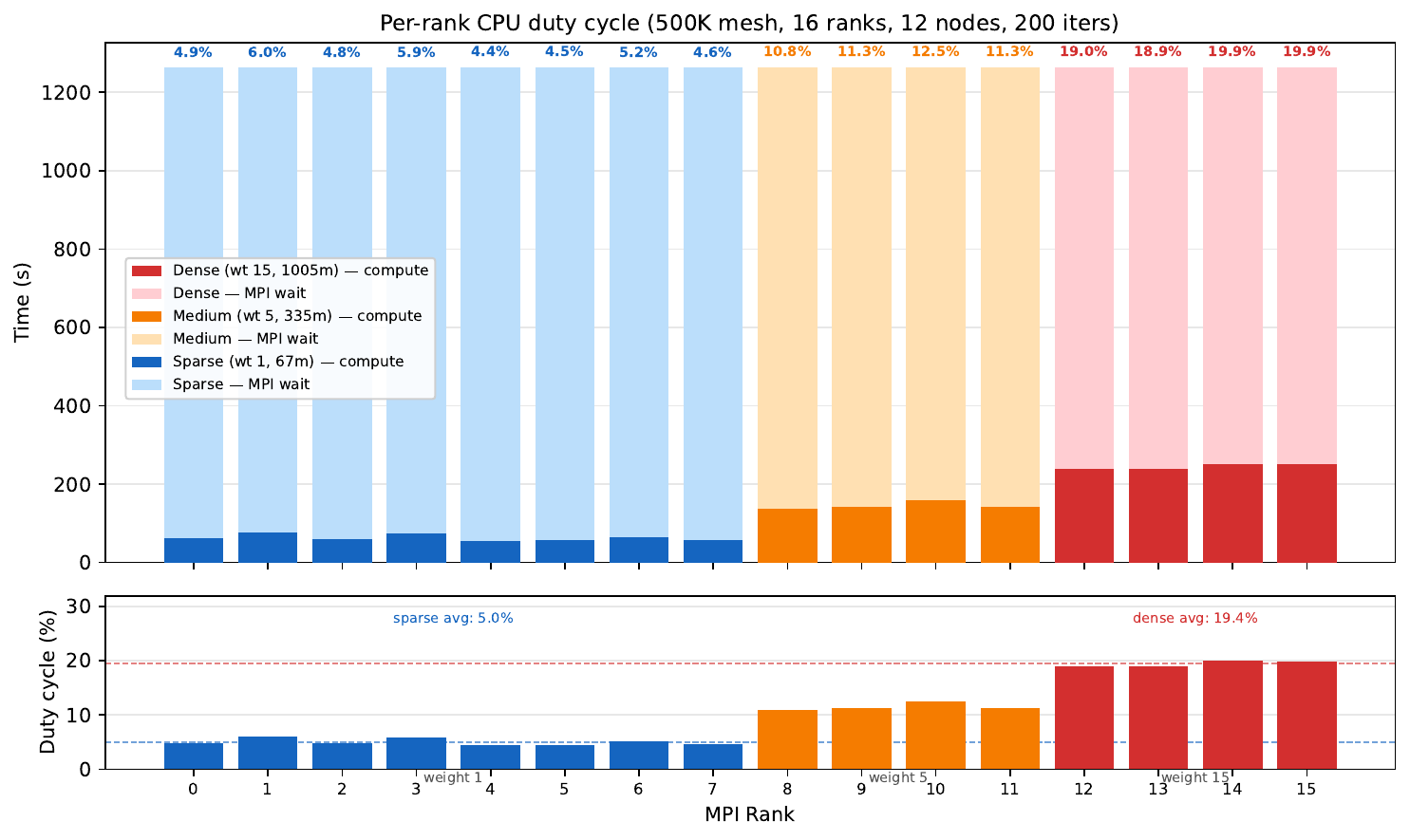}
  \caption{Per-rank CPU duty cycle (498\,834-cell mesh, 16~ranks,
    200~iterations, 12-node cluster, concentric decomposition).
    Top: absolute time split between computation (solid) and MPI
    wait (light).  Bottom: duty cycle percentage.  Sparse ranks
    (blue, weight~1) average 5.0\%, dense ranks (red, weight~15)
    average 19.4\%.}
  \label{fig:duty-cycle}
\end{figure}

Figure~\ref{fig:duty-cycle} presents the per-rank duty cycle measured
on the 498\,834-cell mesh with 16~ranks over 200~solver iterations
across the 12-node homogeneous cluster.
All ranks exhibit low duty cycles, confirming that the majority of
each iteration is spent in MPI synchronization rather than computation.
Sparse ranks (weight~1, ranks~0--7) average 5.0\% duty cycle,
spending 95\% of their time idling at MPI barriers.
Medium ranks (weight~5, ranks~8--11) average 11.5\%, and dense ranks
(weight~15, ranks~12--15) average 19.4\%.

The 3.9$\times$ ratio between dense and sparse duty cycles confirms
that the concentric decomposition creates meaningful load imbalance:
dense ranks perform approximately four times more computation per
iteration than sparse ranks.  However, even the densest ranks spend
80.6\% of their time in MPI communication, indicating that the
498\,834-cell mesh distributed across 16~ranks produces subdomains
small enough that MPI collective overhead dominates per-iteration
wall-clock time for all ranks.

The aggregate reclaimable capacity across all 16~ranks, computed via
Equation~\ref{eq:reclaimable}, represents 88\% of the total CPU
budget.  Profiling runs on the 250K and 1M mesh variants reproduce
the same sparse-to-dense structure; their duty cycles enter the
analysis only through the fitted contention coefficient, and
Section~\ref{sec:beta-sensitivity} shows that the resulting
throughput behavior is nearly mesh-independent.

\begin{figure}[t]
  \centering
  \includegraphics[width=\linewidth]{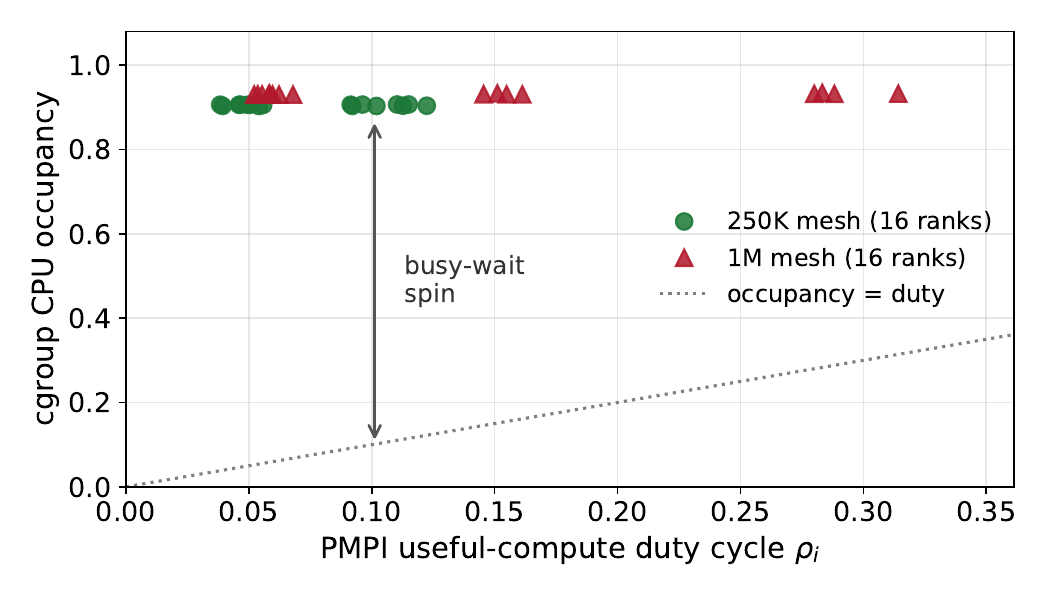}
  \caption{PMPI useful-compute duty cycle versus cgroup CPU occupancy
    per rank (250K and 1M profiling runs, 16~ranks each).  Occupancy
    approaches a full core regardless of duty because the Open~MPI
    progress engine busy-polls at barriers; the vertical gap is spin
    time.}
  \label{fig:duty-validation}
\end{figure}

Figure~\ref{fig:duty-validation} compares, rank by rank, the PMPI
duty cycle against the cgroup CPU occupancy of
Section~\ref{sec:cgroup-method} on the 250K and 1M profiling runs.
Occupancy sits far above duty for every rank and approaches a full
core even for ranks whose useful-compute duty is below 10\%.  This
gap is the busy-wait spin of the MPI progress engine, and it carries
two lessons.  First, CPU-usage counters alone cannot reveal
reclaimable idleness, which is why PMPI (or an equivalent
application-level signal) is required for allocation decisions.
Second, occupancy, not duty, is what saturates cores, which is why
the capacity boundary of Equation~\ref{eq:capacity-boundary} is set
by pod count rather than by aggregate duty cycle.

\subsection{Factorial Comparison}
\label{sec:concurrent-results}

Table~\ref{tab:results} summarizes the wall-clock time for each
configuration, averaged over three repetitions.  For dual-simulation
configurations, the wall clock is defined as the completion time of the
slower simulation (i.e., the makespan).

\begin{table}[t]
  \centering
  \caption{Concurrent simulation results on the 12-node homogeneous cluster:
    wall-clock time, throughput gain, and fairness
    (mean $\pm$ std over 3~runs, 200~solver iterations each).}
  \label{tab:results}
  \footnotesize
  \begin{tabular}{@{}lrrrrr@{}}
    \toprule
    \textbf{Config} & \textbf{Sim\,A (s)} & \textbf{Sim\,B (s)} &
    \textbf{Makespan (s)} & \textbf{Thpt.} & \textbf{A/B} \\
    \midrule
    C-1E & $1179 \pm 28$ & --- & 1179 & 1.00$\times$ & --- \\
    C-1P & $1249 \pm 20$ & --- & 1249 & --- & --- \\
    C-2P & $1331 \pm 12$ & $1410 \pm 10$ & 1410 & \textbf{1.77$\times$} & 0.94 \\
    C-2E & $1286 \pm 14$ & $1274 \pm 22$ & 1286 & 1.83$\times$ & 1.01 \\
    \bottomrule
  \end{tabular}
\end{table}

Both dual-simulation configurations
achieve substantial throughput gains.
Configuration~C-2P completes
two simulations in \SI{1410}{\second}, compared to
$2 \times \SI{1249}{\second} = \SI{2498}{\second}$ for sequential C-1P,
yielding a throughput gain of $1.77\times$.  Configuration~C-2E achieves
$1.83\times$ throughput by completing both simulations in
\SI{1286}{\second}.  Running two simulations concurrently adds only
9--13\% to the wall-clock time of a single simulation.
C-2E achieves slightly higher raw throughput than C-2P, but consumes
a much larger share of the cluster's schedulable request budget
(\SI{32}{vCPU} vs.\ \SI{12}{vCPU}), which limits co-location density
at higher~$N$ (Section~\ref{sec:pareto}).

Turning to fairness, both allocation strategies produce
co-location ratios within 10\% at $N=2$: C-2P has an A/B
wall-clock ratio of 0.94 (Sim~A: $1331 \pm 12$\,s, Sim~B:
$1410 \pm 10$\,s) and C-2E has 1.01.  We note that in C-2P,
Sim~B is systematically 6\% slower than Sim~A across all three
repetitions, likely due to pod scheduling order (Sim~B's pods are
deployed second and may land on nodes already hosting Sim~A's
dense ranks, increasing contention).  The same signature recurs at
every higher density in Section~\ref{sec:pareto}: later-deployed
simulations run marginally longer, yet the slowest-to-fastest ratio
never exceeds 1.08 in any of the concurrent runs.  The effect is
systematic but small; the absolute difference ($\sim$79\,s) is
minor relative to the makespan, and both simulations complete
within the same scheduling window.

At higher co-location densities, per-node contention increases and
fairness may diverge between allocation strategies.
Section~\ref{sec:pareto} tests this by increasing the number of
concurrent simulations through $N=8$, past the capacity boundary.

Comparing the single-simulation baselines, C-1P is 6\% slower than C-1E
(\SI{1249}{\second} vs.\ \SI{1179}{\second}), and C-2E outperforms
C-2P by 9\% in raw makespan (\SI{1286}{\second} vs.\
\SI{1410}{\second}).  Under proportional allocation, sparse ranks
request only \SI{67}{\milli\cpu}, giving them lower CFS weight
and reduced burst priority compared to equal allocation's
\SI{1000}{\milli\cpu}.  On the lightly loaded 12-node cluster,
this translates to slightly less CPU time during burst.

However, the value of proportional allocation lies not in raw speed
but in \emph{resource budget efficiency}.  Equal allocation at $N=5$
requires $5 \times 16 \times \SI{1000}{\milli\cpu} =
\SI{80}{vCPU}$ of aggregate requests, consuming 83\% of the cluster.
Proportional allocation requires only $5 \times \SI{5900}{\milli\cpu}
\approx \SI{30}{vCPU}$, consuming 31\%.  On a shared cluster with a
Kubernetes \texttt{ResourceQuota}, proportional allocation enables
more concurrent simulations within the same quota.  The Pareto sweep
in Section~\ref{sec:pareto} demonstrates this: C-5P fits comfortably
at \SI{30}{vCPU}, whereas C-5E would require \SI{80}{vCPU} and may
exceed a team's quota allocation.

\begin{figure}[t]
  \centering
  \includegraphics[width=\linewidth]{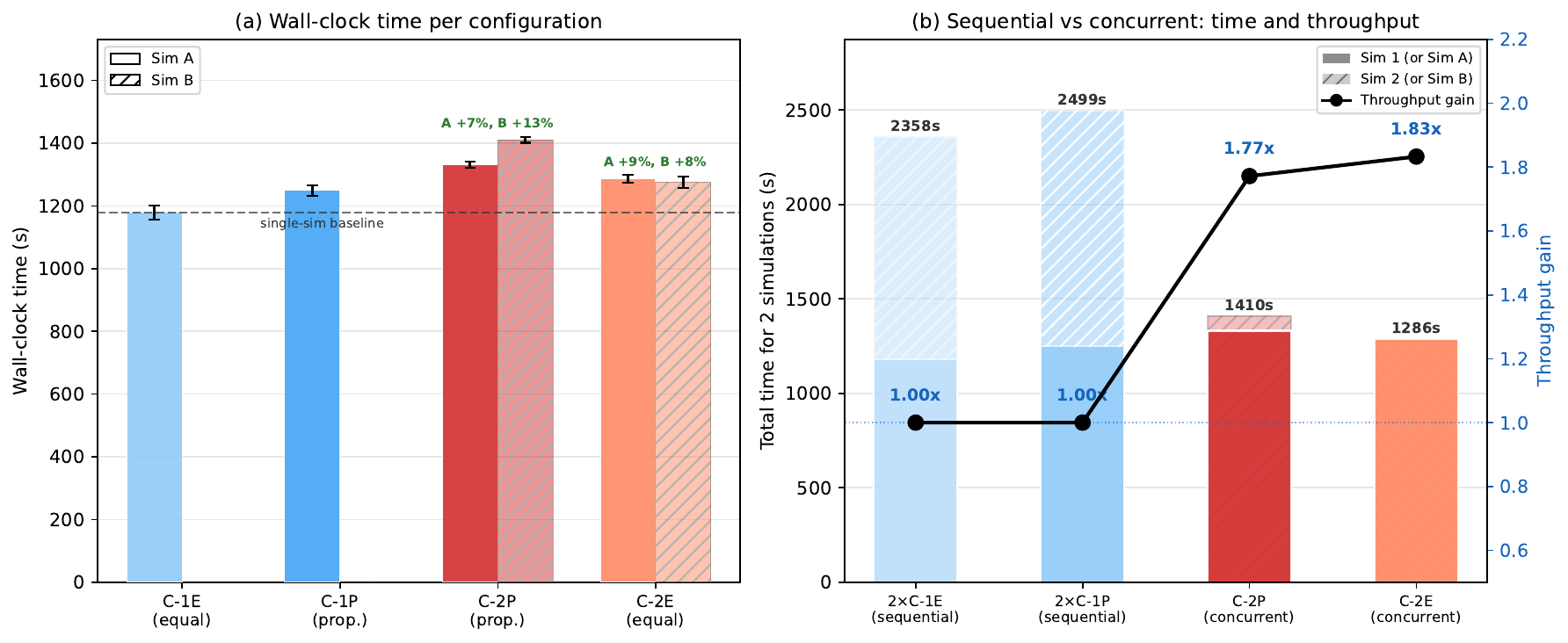}
  \caption{(a)~Wall-clock time per configuration (solid: Sim~A,
    hatched: Sim~B).
    (b)~Total time to complete two simulations: sequential
    ($2\times$ single-sim, stacked) vs.\ concurrent (overlapping).
    Black line shows throughput gain (right axis).
    Concurrent execution nearly halves total time:
    C-2P $1.77\times$, C-2E $1.83\times$.}
  \label{fig:throughput}
\end{figure}

Figure~\ref{fig:throughput} visualizes these results: the left panel
shows per-configuration wall-clock times with degradation annotations,
and the right panel compares sequential versus concurrent total time.

\subsection{Pareto Sweep to the Capacity Boundary}
\label{sec:pareto}

\begin{figure}[t]
  \centering
  \includegraphics[width=\linewidth]{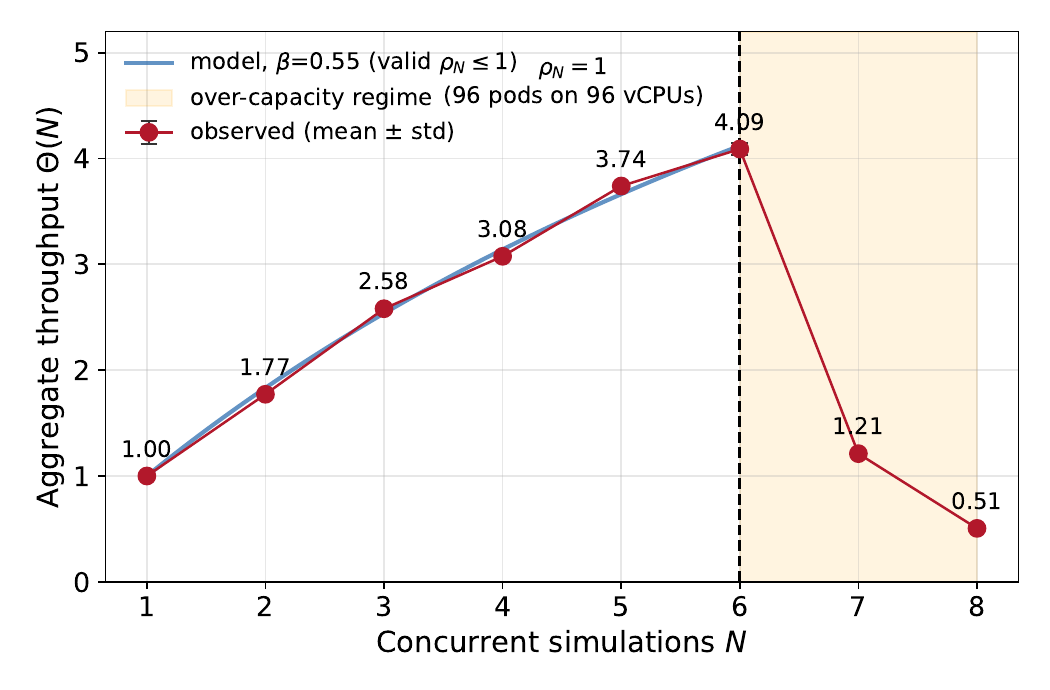}
  \caption{Throughput versus concurrency (mean $\pm$ standard
    deviation over runs) with the single-parameter model
    ($\beta = 0.55$, fitted over the valid domain).  The model
    tracks observation within 3.4\% for $\rho_N \le 1$ and is not
    extended beyond that domain.  Past the boundary (shaded, 112
    pods on 96 vCPUs at $N=7$), busy-wait spin displaces
    computation and throughput collapses to $0.50\times$ at $N=8$.}
  \label{fig:pareto-cliff}
\end{figure}

To identify the optimal co-location density and to locate the limit of
the approach, we extend the proportional-allocation experiments to
$N = 3, 4, 5, 6, 7$, and~$8$ concurrent simulations.
Figure~\ref{fig:pareto-cliff} plots the resulting throughput
(mean $\pm$ standard deviation over runs) together with the model.

Below the capacity boundary, throughput scales sub-linearly but
remains substantial: $1.77\times$ at $N=2$, $2.58\times$ at $N=3$,
$3.08\times$ at $N=4$, $3.74\times$ at $N=5$, and $4.09\times$ at
$N=6$.  The corresponding scheduling efficiency (throughput$/N$)
decreases from 89\% to 68\%, with the steepest early drop between
$N=3$ (86\%) and $N=4$ (77\%).  This identifies $N=3$ as the knee of
the Pareto curve for latency-sensitive teams, offering the best
trade-off between throughput and per-case degradation (16\%, versus
47\% at $N=6$).  Teams optimizing pure throughput can push to
$N = N_{\max} = 6$.

Crossing the boundary changes the character of the system.  At $N=7$
(112 worker pods on 96~vCPUs) the mean makespan jumps from
1833\,s to 7221\,s and throughput collapses to $1.21\times$; at
$N=8$ it falls to $0.50\times$, meaning eight concurrent simulations
complete \emph{half} as much work per hour as running them one at a
time.  Mean iteration time grows from 9\,s to 36\,s to 99\,s across
$N = 6, 7, 8$.  The collapse is stable and reproducible (run-to-run
variation under 2\% in the over-capacity regime) and symmetric
(fairness ratio at most 1.06), ruling out transient interference or
a single straggler.  Node CPU utilization is pegged near 95\% on
both sides of the boundary and no CFS throttling occurs (no limits
are set), confirming the mechanism identified in
Section~\ref{sec:theory}: once runnable threads exceed cores,
busy-wait spin competes with computation on every core, and every
barrier inflates.  The penalty for exceeding capacity is not
graceful degradation but outright loss; a practitioner who
oversubscribes by one simulation forfeits the entire multiplexing
gain and more.

The model of Equation~\ref{eq:throughput-model} captures the entire
valid domain with a single parameter.  Fitting $\beta$ over
$N = 2\ldots6$ yields $\beta = 0.55$, which reproduces every 500K
measurement within 3.4\% and the boundary point $N=6$ within 0.8\%
(predicted $4.06\times$, observed $4.09\times$).  Calibrating from
the $N=2$ point alone gives $\beta = 0.77$ and systematically
conservative predictions that under-estimate throughput by 7\% at
$N=3$ and by at most 11\% at $N=6$; for capacity planning this bias
errs on the safe side.  The model makes no
prediction beyond $\rho_N = 1$, and the data confirm that none
should be attempted: the first over-capacity point lies $73$\% below
the (invalid) linear extrapolation of the domain-wide fit.

\subsection{Sensitivity of the Model Across Mesh Sizes}
\label{sec:beta-sensitivity}

A single fitted $\beta$ is only useful if it transfers.  We repeat
the profiling run and configurations C-1P, C-2P, and C-3P on the
250K and 1M meshes and ask two questions: does the throughput gain
itself depend on workload intensity, and does a $\beta$ fitted on one
mesh predict the others?

\begin{figure}[t]
  \centering
  \includegraphics[width=\linewidth]{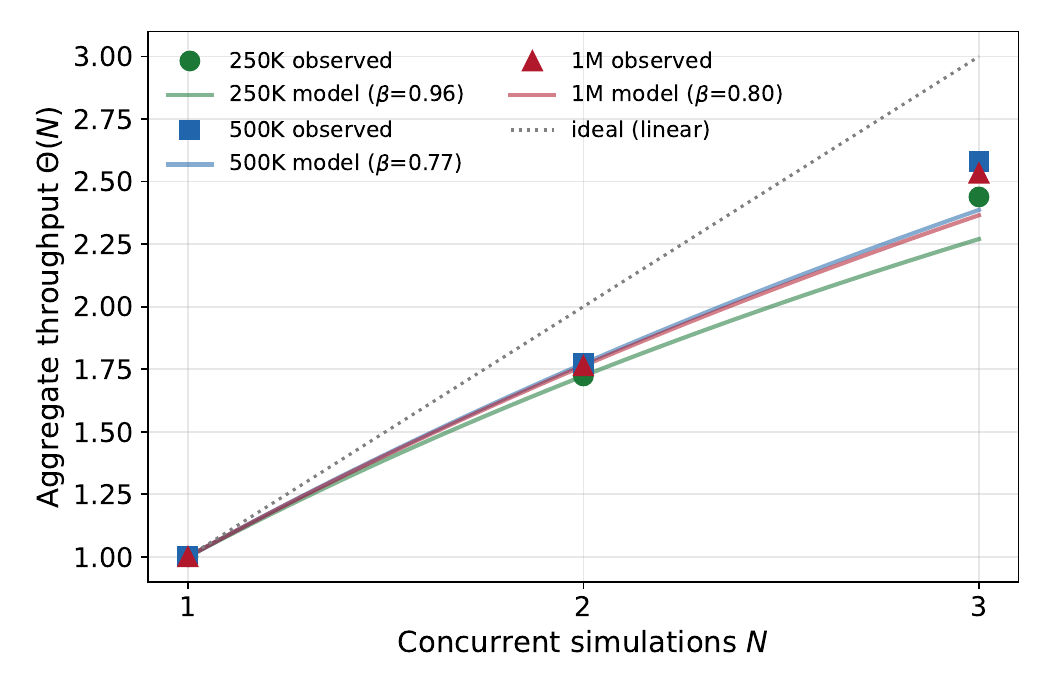}
  \caption{Throughput versus concurrency for three mesh sizes
    spanning $4\times$ in cell count.  Markers: observation; lines:
    per-mesh model fits ($\beta = 0.96/0.77/0.81$).  The gains are
    nearly mesh-independent, and the 500K-fitted $\beta$ predicts
    the other meshes within 5.8\%.}
  \label{fig:beta-sensitivity}
\end{figure}

The gains are strikingly stable.  Across a $4\times$ range of mesh
size, $\Theta(2) = 1.72 / 1.77 / 1.76$ and
$\Theta(3) = 2.44 / 2.58 / 2.53$ for the 250K/500K/1M meshes
respectively.  Fitting $\beta$ per mesh from each $N=2$ point yields
$0.96 / 0.77 / 0.81$, an $11$\% spread around the mean.  Most
practically relevant, applying the 500K-fitted $\beta = 0.77$
unchanged to the other meshes predicts all four cross-mesh
configurations within $5.8$\%, and the domain-wide
$\beta = 0.55$ of Section~\ref{sec:pareto} reproduces all twelve
in-domain configurations, across every mesh and concurrency, within
$6.4$\% (Figure~\ref{fig:beta-sensitivity}).
The single-parameter model is therefore not a curve-fitting artifact
of one workload: within this workload class, one calibration on one
mesh suffices for capacity planning across substantially
different mesh resolutions.  The residual mesh dependence is
consistent with the duty-cycle differences of
Section~\ref{sec:duty-results}: the 250K mesh has the lowest duty
cycles and the highest per-iteration fixed overheads, and
accordingly shows the largest single deviation (6.4\%, at $N=2$)
under the transferred coefficient.

\subsection{Generality: NAS Parallel Benchmarks}
\label{sec:npb-results}

\begin{figure}[t]
  \centering
  \includegraphics[width=\linewidth]{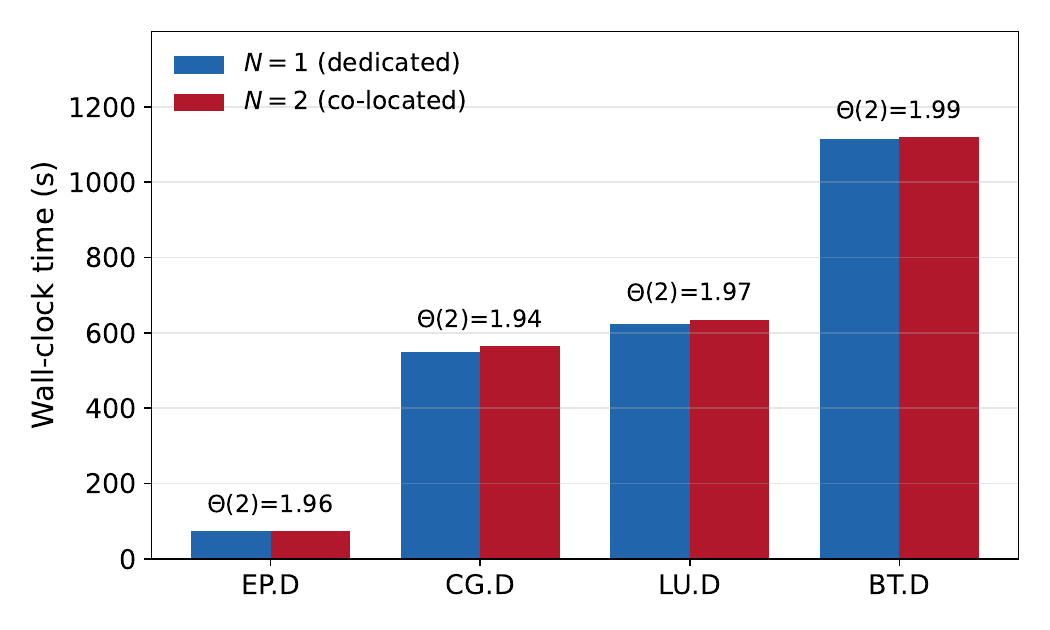}
  \caption{NAS Parallel Benchmarks (class D, 16 ranks): dedicated
    ($N=1$) versus co-located ($N=2$) wall-clock time, three runs
    each.  All four kernels retain $\Theta(2) \ge 1.92$.}
  \label{fig:npb}
\end{figure}

Figure~\ref{fig:npb} reports the NPB results.  All four kernels
co-locate at $N=2$ nearly for free: $\Theta(2) = 1.92$ (EP), $1.92$
(CG), $1.96$ (LU), and $1.99$ (BT), with makespan overheads of
0.5--4.4\% and all verification checks passing.  The cgroup occupancy
of the kernels spans 0.75--0.98, so this is not because the kernels
are idle; it is because $2 \times 16$ busy ranks on 96~vCPUs leave
$\rho_N$ well below~1, exactly the regime in which the capacity
analysis predicts near-zero interference.  The result generalizes the
central claim beyond CFD: co-location on Kubernetes with
requests-only allocation is essentially free while the pod count
stays within the vCPU count, for workloads ranging from
embarrassingly parallel (EP) to communication-intensive (CG).  It
also delineates where sparse-domain allocation specifically matters:
load-balanced kernels gain nothing from proportional requests, whereas
imbalanced CFD decompositions release five sixths of their request
budget (Section~\ref{sec:concurrent-results}), which is what allows
$N$ to grow under a namespace quota.

\subsection{Sensitivity to Cluster Size}
\label{sec:nodescale-results}

Repeating C-1P and C-2P with six of the twelve workers cordoned
tests whether the observed gains depend on the specific
twelve-node geometry.  They do not.  On six nodes the C-1P baseline
completes in $1269 \pm 6$\,s (within 1.6\% of the twelve-node
$1249$\,s despite double the per-node pod density), and dual
simulation achieves $\Theta(2) = 1.68$ versus $1.77$ on twelve
nodes.  The reduction is consistent with the model: halving the
cluster doubles the load increment per simulation, moving $N=2$ on
six nodes to the same $\rho_N$ as $N=4$ on twelve, where efficiency
is likewise 77\%.  The mechanism,
and its predictability, carry over to the smaller cluster that many
practitioner teams would actually operate.

\subsection{Memory Footprint}
\label{sec:memory-results}

\begin{figure}[t]
  \centering
  \includegraphics[width=\linewidth]{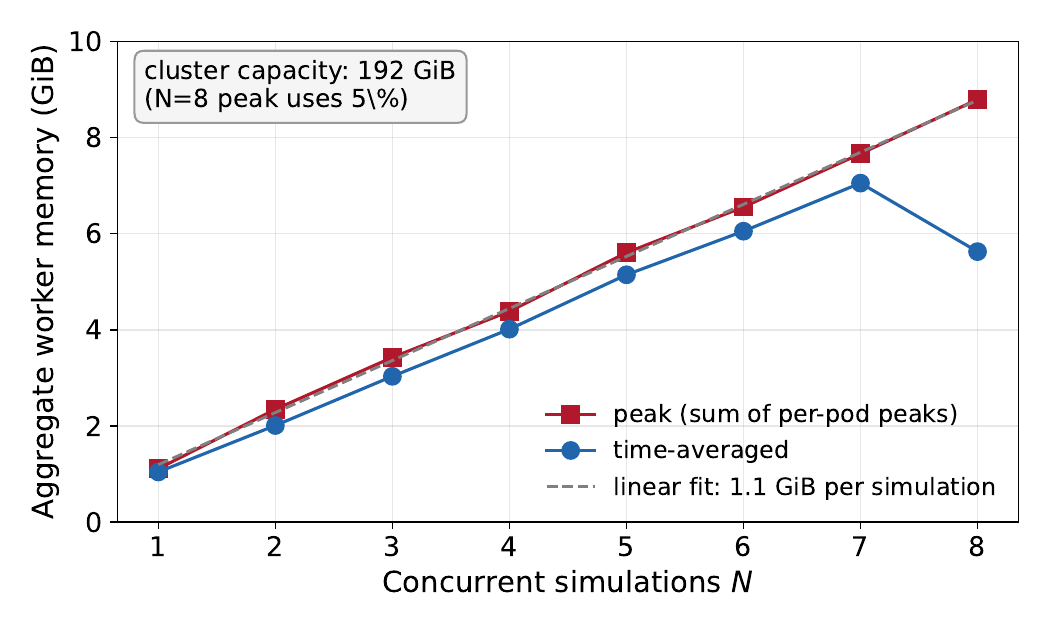}
  \caption{Aggregate worker memory versus concurrency (mean of
    per-run values; peak = sum of per-pod maxima).  Growth is linear
    at 1.08\,GiB per simulation and reaches only 5\% of cluster
    memory at $N=8$.}
  \label{fig:memory}
\end{figure}

CPU is the multiplexed resource, but co-location is only safe if
memory does not become the binding constraint first.
Figure~\ref{fig:memory} reports per-pod memory sampled every 5\,s and
aggregated per configuration.  Aggregate worker memory grows linearly
at 1.08\,GiB per additional simulation (peak-of-run sums; per-pod
peaks range from about 38\,MiB for sparse ranks to 138\,MiB for
dense ranks).  Even at $N=8$ (128 worker pods) the aggregate peak
of 8.8\,GiB is under 5\% of the cluster's 192\,GiB.  For this
workload class, memory is far from limiting at every achievable
concurrency, and the capacity boundary of
Section~\ref{sec:pareto} is reached long before memory pressure
could appear.  Larger meshes shift this balance (the 1M mesh roughly
doubles per-rank memory), and memory-bandwidth contention, which we
do not measure directly, remains the caveat identified in
Section~\ref{sec:related-work}.

\subsection{Dynamic Controller Evaluation}
\label{sec:dynamic-eval}
To demonstrate that the multiplexing framework can be fully
automated, we evaluate the controller of
Section~\ref{sec:controller-design} end to end.  Starting from a
single simulation with equal CPU allocation
(\SI{1000}{\milli\cpu} per rank), the controller executed the full
Profile--Resize--Pack--Monitor pipeline without manual
intervention.

\begin{figure}[t]
  \centering
  \includegraphics[width=\linewidth]{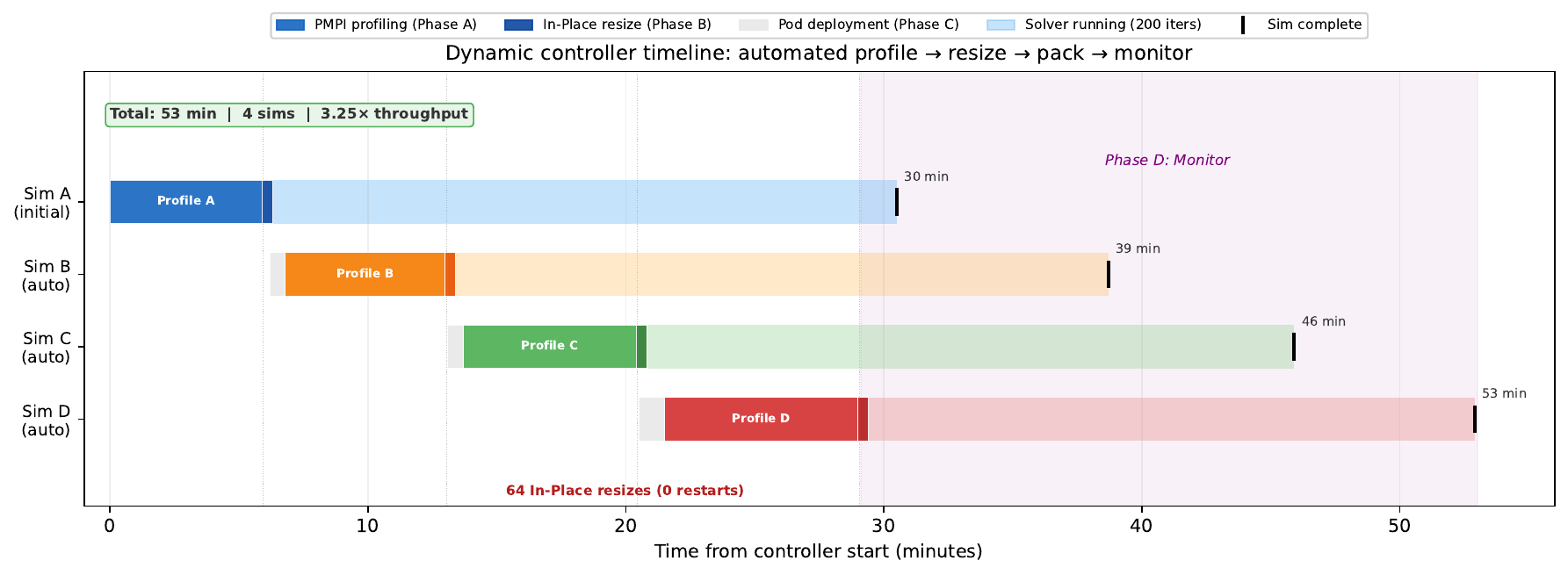}
  \caption{Dynamic controller timeline.  The controller automatically
    profiles each simulation via PMPI (dark bars), resizes all pods
    via KEP-1287 (thin dark bars), deploys additional simulations
    (grey bars), and monitors fairness.  Lighter bars show solver
    execution (200~iterations).  Total pipeline: 53~minutes from a
    single initial simulation to 4~concurrent simulations at
    $3.25\times$ throughput with 64~in-place resizes and zero pod
    restarts.}
  \label{fig:dynamic-timeline}
\end{figure}

In an end-to-end test, the controller automatically scaled from 1 to 4
concurrent simulations in 53~minutes.  It performed 4~PMPI profiling
passes, 64~in-place pod resizes, 3~automatic simulation deployments
(including mesh decomposition and MPI launch), and 2~fairness
adjustments.  The resulting makespan of \SI{1537}{\second} for 4
simulations corresponds to $3.25\times$ throughput, exceeding the
static C-4P result ($3.08\times$) by 6\% and validating that automated
duty-cycle-aware allocation matches or improves upon manual
configuration.  To our knowledge, this is the first application of
in-place CPU scaling (KEP-1287) to running MPI
workloads~\cite{kep1287}; prior work (ARC-V~\cite{medeiros2025arcv})
applied in-place scaling to memory only.
Figure~\ref{fig:dynamic-timeline} illustrates the end-to-end timeline.

We emphasize that this is a \emph{proof-of-concept} demonstration
based on a single end-to-end run, not a production-hardened system.
The controller does not currently handle failure modes such as
kubelet resource pressure, resize request rejection, or partial
simulation deployment.  Repeated runs, fault-injection testing,
and formal robustness evaluation are planned as future work.
The controller is composable with cluster-level schedulers such as
Volcano and Kueue, which handle inter-job queue management; our
controller operates at the intra-job rank level and is orthogonal
to these tools.

\section{Discussion}
\label{sec:discussion}

\paragraph{Why multiplexing works, and where it stops}
The multiplexing mechanism relies on CFS time-slicing between
concurrent simulations' MPI barrier phases.  Because each
simulation has an independent \texttt{MPI\_COMM\_WORLD}, their
compute and wait phases are unsynchronized and naturally interleave:
when one simulation's ranks idle at a barrier, the kernel
context-switches their CPU cycles to the other simulation within
microseconds.  The low duty cycles (5--20\%) ensure that compute
phase overlaps are rare ($P(\geq 2) \approx 0.19$ at $N=5$,
see below), limiting contention.  The same busy-wait behavior that
makes barrier cycles preemptible also sets the hard boundary of the
approach: a spinning rank yields its core only through preemption,
so the mechanism works exactly as long as every runnable thread can
be granted a core when it matters ($\rho_N \le 1$).  An alternative
worth noting is Open~MPI's \texttt{mpi\_yield\_when\_idle} option,
which makes waiting ranks yield voluntarily; it would soften the
over-capacity collapse at the price of barrier-exit latency for all
regimes, and evaluating this trade-off is left to future work.

\paragraph{How the gains compare with prior co-location results}
The magnitudes observed here sit within, and help explain, the range
reported by the co-location literature.  Iancu
et~al.~\cite{iancu2010oversubscription} found oversubscription
gains of up to $1.3$--$2\times$ for workloads with complementary
demands, and sharp degradation for synchronization-heavy codes;
our results reproduce both regimes and separate them with an
explicit, predictable boundary.  Heracles~\cite{lo2015heracles}
sustained about 90\% machine utilization when co-locating
latency-critical and batch workloads using hardware partitioning;
our approach reaches its comparable operating point (95\% node CPU
at $N=6$, $69$\% scheduling efficiency) with no hardware isolation
features, using only declarative Kubernetes requests.  Gang
scheduling~\cite{feitelson1992gang}, and its Kubernetes descendants
Volcano and Kueue, would execute our workload sequentially, which is
precisely the $1.00\times$ baseline all our gains are measured
against.  ARC-V~\cite{medeiros2025arcv}, the only prior use of
in-place pod resizing for HPC, targets memory and does not co-locate
CPU; the $1.77$--$4.09\times$ range reported here is, to our
knowledge, the first quantified CPU-multiplexing envelope for
tightly-coupled MPI jobs on Kubernetes.

\paragraph{Practical guidelines and alternatives}
For practitioners deciding whether to enable multiplexing, we
recommend the following workflow: (1)~run a short PMPI profiling pass
(50~iterations, approximately 5~minutes) to measure per-rank duty
cycles; (2)~if the average duty cycle is below 50\%, substantial
idle capacity exists and multiplexing is beneficial; (3)~use the
throughput model (Equation~\ref{eq:throughput-model}) to estimate
the optimal~$N$; (4)~deploy additional simulations with proportional
CPU requests.  The dynamic controller automates steps~(1)--(4).
Alternative approaches include spot instances (60--90\% cost
reduction but with preemption risk), managed HPC services such as
AWS ParallelCluster (SLURM-based, whole-node allocation), and
reducing the rank count to raise the per-rank duty cycle at the expense
of co-location capacity.  The optimal strategy depends on a team's
cost, latency, and availability constraints.

\begin{table}[t]
  \centering
  \caption{Cost per simulation at different co-location densities
    (12-node c5.2xlarge cluster, \$4.12/hr on-demand).}
  \label{tab:cost}
  \small
  \begin{tabular}{@{}rrrrc@{}}
    \toprule
    $N$ & Total time (s) & Total cost (\$) & \$/sim & Saving \\
    \midrule
    1 & 1249 & 1.43 & 1.43 & baseline \\
    2 & 1410 & 1.61 & 0.81 & 44\% \\
    3 & 1452 & 1.66 & 0.55 & 62\% \\
    4 & 1625 & 1.86 & 0.46 & 68\% \\
    5 & 1670 & 1.91 & 0.38 & 73\% \\
    6 & 1833 & 2.10 & 0.35 & 76\% \\
    \bottomrule
  \end{tabular}
\end{table}

\paragraph{Decomposition independence}
Most CFD practitioners use default equal-weight graph partitioners
(Scotch, METIS) without considering domain-specific load imbalance.
The concentric decomposition used in this study makes the physical
near-wall/far-field imbalance explicit.  To verify that multiplexing
does not depend on this choice, we repeated the single-simulation PMPI profiling with
standard Scotch equal partitioning: all 16~ranks exhibited duty
cycles of 7.7--13.1\% (87--92\% idle), and wall-clock time was
within 1\% of the concentric result (\SI{1272}{\second} vs.\
\SI{1263}{\second}).  Multiplexing benefits
therefore persist under any partitioning strategy.

\paragraph{Scope and limitations}
Profiling with \texttt{hostNetwork} pods showed essentially unchanged
duty-cycle ratios (sparse 4.7--5.8\% vs.\ 4.4--6.0\% with pod
networking, dense 19.8--22.7\% vs.\ 18.9--19.9\%), confirming that
the load imbalance is inherent to the MPI protocol, not a
container-networking artifact.  Memory-bandwidth contention is not the limiting factor
at the duty cycles observed.  On a node hosting $K$ pods each with
duty cycle $d$, the probability that two or more pods compute
simultaneously (and thus compete for memory bandwidth) is
$P(\geq 2) = 1 - (1-d)^K - Kd(1-d)^{K-1}$.  At $N=5$
($K \approx 6.7$, $d \approx 0.12$), this gives $P \approx 0.19$,
consistent with the observed 10--13\% per-case degradation and
suggesting that rare compute-phase overlaps account for most of
the multiplexing overhead.  Direct validation through hardware
performance-monitoring counters (last-level-cache miss rates) is
not possible on Nitro-virtualized c5.2xlarge instances, which do
not expose these counters to guest VMs.

The experiments use one CFD solver (OpenFOAM
\texttt{rhoSimpleFoam}) on one geometry (NACA~0012) at three mesh
resolutions, four NPB kernels, one rank count (16), two cluster
sizes (twelve and six nodes), and single-NUMA-domain instances.
For compute-intensive workloads with higher duty cycles (e.g.,
detailed chemical kinetics), NUMA-aware scheduling and intentional
sparse-dense pod pairing may become relevant.  The optimal rank
count is an engineering trade-off between single-case speed and
co-location capacity.  Fault tolerance is out of scope by design:
the framework adds no failure modes beyond stock Kubernetes (pods
are ordinary pods, and a failed simulation affects others only by
releasing its CPU weight), and across the 175 CFD and 36 NPB
executions reported
here no run failed; but node-loss recovery for the MPI jobs
themselves remains the responsibility of the application layer, as
in all the Kubernetes-MPI work cited in
Section~\ref{sec:related-work}.

\section{Conclusions and Future Work}
\label{sec:conclusions}

\subsection{Conclusions}

This paper demonstrated that MPI-parallel CFD simulations on
Kubernetes waste 80--95\% of their provisioned CPU cycles at
synchronization barriers, and that this idle capacity can be
reclaimed by co-locating concurrent simulations on the same cluster.

Six principal findings emerged.  First, PMPI profiling revealed
that all ranks, including dense near-wall ranks, spend 80--95\% of
their time at MPI barriers, with this behavior persisting under both
concentric and standard Scotch partitioning.  Second, co-locating
two simulations adds only 9--13\% overhead, yielding $1.77\times$
throughput under proportional allocation and 44\% cost reduction.
Third, throughput scales to $4.09\times$ at $N = N_{\max} = 6$ with
a Pareto knee at $N=3$ (86\% efficiency), and a single contention
coefficient fitted on one mesh reproduces all twelve in-domain
configurations across a $4\times$ mesh-size range within 6.4\%,
including the boundary point within 0.8\%.  Fourth, the capacity
boundary is real, sharp, and predictable: beyond $\rho_N = 1$,
busy-wait MPI spin displaces computation and throughput collapses
to $0.50\times$ at $N=8$, worse than sequential execution.  Fifth,
the mechanism generalizes: four NPB kernels co-locate at
$1.92$--$1.99\times$, a half-size cluster reproduces the gains, and
memory grows linearly at 1.1\,GiB per simulation, never
approaching cluster capacity.  Sixth, a dynamic controller automated
the full pipeline via KEP-1287 in-place CPU scaling, achieving
$3.25\times$ for four simulations with zero pod restarts.

These results demonstrate that Kubernetes, despite its microservices
heritage, can serve as a cost-effective platform for throughput-oriented
engineering simulation workloads.

\subsection{Future Work}

Two extensions are planned.  First, \emph{heterogeneous workload
pairing}: co-locating simulations with complementary computational
profiles (e.g., a fine-mesh turbulence simulation with a coarse-mesh
screening run) using Kubernetes pod affinity rules to maximize
CPU utilization through intentional sparse-dense rank co-location.

Second, \emph{combustion chemistry acceleration}: in reactive CFD,
detailed chemical kinetics creates extreme load imbalance (reacting
cells require 100--500$\times$ more sub-iterations than non-reacting
cells)~\cite{xie2024flamelet}.  Combining a neural network surrogate
for non-stiff cells with the proportional scheduling framework
presented here could simultaneously reduce per-rank compute cost and
improve cluster packing efficiency.

\subsection*{CRediT authorship contribution statement}

\textbf{Tianfang Xie:} Conceptualization, Methodology, Software,
Validation, Formal analysis, Investigation, Data curation, Writing
-- original draft, Writing -- review \& editing, Visualization.

\subsection*{Declaration of competing interest}

The author declares that he has no known competing financial
interests or personal relationships that could have appeared to
influence the work reported in this paper.

\subsection*{Data availability}

All experiment scripts, Kubernetes manifests, the PMPI profiling
library, the dynamic controller, analysis code, and the measurement
data underlying every figure are publicly available at
\url{https://github.com/Xieldor/K8s-CFD-Multiplexing}.

\bibliographystyle{elsarticle-num}
\bibliography{references}

\end{document}